\documentclass[onecolumn]{aastex631}

\newcommand{\Msun}{$M_{\odot}$}
\newcommand{\Rsun}{$R_{\odot}$}

\newcommand{\Mjup}{M$_{\mathrm{Jup}}$}
\newcommand{\Teff}{$T_{\mathrm{eff}}$}

\newcommand{\logg}{$\log{g}$}
\newcommand{\Rjup}{R$_{\mathrm{Jup}}$}

\begin{document}

\title{Metal Polluted White Dwarfs with 21~\micron\ IR excesses from JWST/MIRI: Planets or Dust?}

\author[0000-0002-1783-8817]{John H. Debes}
\affiliation{The Space Telescope Science Institute\\
3700 San Martin Dr. \\
Baltimore, MD 21218, USA}
\affiliation{Aura, for ESA}

\author[0009-0008-7425-8609]{Sabrina Poulsen}
\affiliation{Homer L. Dodge Department of Physics and Astronomy, University of Oklahoma, 440 W. Brooks St., Norman, OK, 73019 USA}

\author[0009-0002-4970-3930]{Ashley Messier}
\affiliation{Five College Astronomy Department, Amherst College, Amherst, MA 01002, USA}

\author[0000-0001-7106-4683]{Susan E. Mullally}
\affiliation{The Space Telescope Science Institute\\
3700 San Martin Dr. \\
Baltimore, MD 21218, USA}

\author[0009-0004-6806-1675]{Katherine Thibault}
\affiliation{D\'epartement de physique and Institut Trottier de recherche sur les exoplan\`etes, Universit\'e de Montr\'eal, C.P. 6128, Succ. Centre-ville, Montr\'eal, H3C 3J7, Québec, Canada}
\affiliation{Observatoire du Mont-M\'egantic, Universit\'e de Montr\'eal, C.P. 6128, Succ. Centre-ville, Montr\'eal, H3C 3J7, Québec, Canada}

\author[0000-0003-0475-9375]{Lo\"ic Albert}
\affiliation{D\'epartement de physique and Institut Trottier de recherche sur les exoplan\`etes, Universit\'e de Montr\'eal, C.P. 6128, Succ. Centre-ville, Montr\'eal, H3C 3J7, Québec, Canada}
\affiliation{Observatoire du Mont-M\'egantic, Universit\'e de Montr\'eal, C.P. 6128, Succ. Centre-ville, Montr\'eal, H3C 3J7, Québec, Canada}

\author[0000-0002-7698-3002]{Misty Cracraft}
\affiliation{The Space Telescope Science Institute\\
3700 San Martin Dr. \\
Baltimore, MD 21218, USA}

\author[0000-0002-3307-1062]{\'Erika Le Bourdais}
\affiliation{D\'epartement de physique and Institut Trottier de recherche sur les exoplan\`etes, Universit\'e de Montr\'eal, C.P. 6128, Succ. Centre-ville, Montr\'eal, H3C 3J7, Québec, Canada}

\author[0000-0003-4609-4500]{Patrick Dufour}
\affiliation{D\'epartement de physique and Institut Trottier de recherche sur les exoplan\`etes, Universit\'e de Montr\'eal, C.P. 6128, Succ. Centre-ville, Montr\'eal, H3C 3J7, Québec, Canada}

\author[0000-0001-7139-2724]{Tom Barclay}
\affiliation{NASA Goddard Space Flight Center, Greenbelt, MD 20771, USA}
\affiliation{University of Maryland, Baltimore County, Baltimore, MD 21250, USA}

\author[0000-0001-5941-2286]{J. J. Hermes}
\affiliation{Department of Astronomy, Boston University, 725 Commonwealth Avenue, Boston, MA 02215, USA}

\author[0000-0001-6098-2235]{Mukremin Kilic} 
\affiliation{Homer L. Dodge Department of Physics and Astronomy, University of Oklahoma, 440 W. Brooks St., Norman, OK, 73019 USA}

\author[0000-0002-6780-4252]{David Lafreni\`ere}
\affiliation{Institut Trottier de Recherche sur les exoplan\`etes, Universit\'e de Montr\'eal, C.P. 6128, Succ. Centre-ville, Montr\'eal, H3C 3J7, Québec, Canada}
\affiliation{D\'epartement de Physique, Universit\'e de Montr\'eal, C.P. 6128, Succ. Centre-ville, Montr\'eal, H3C 3J7, Québec, Canada}

\author[0009-0004-7656-2402]{Fergal Mullally}
\affiliation{Constellation, 1310 Point Street, Baltimore, MD 21231, USA}

\author[0000-0001-8362-4094]{William Reach}
\affiliation{Space Science Institute, 4765 Walnut Street, Suite 205, Boulder, CO 80301, USA}

\author{Elisa Quintana}
\affiliation{NASA Goddard Space Flight Center, Greenbelt, MD 20771, USA}



\begin{abstract}
White dwarfs with metal pollution are caused by the accretion of rocky dust from tidally disrupted minor bodies and are signposts for planetary systems. These minor bodies are perturbed by planets that have survived post-main sequence evolution. Open questions exist as to the typical mass of the perturbers and the specific planetary architectures that cause metal pollution. JWST's sensitivity in the mid-IR has opened new doors to deciphering polluted white dwarfs.  We present JWST Cycle 1 MIRI imaging of four nearby metal-polluted white dwarfs with the goal of detecting and characterizing planetary companions. With deep mid-IR imaging we are sensitive to cold Jupiter-mass planet analogs. In addition to finding two candidate planetary companions, for the first time we detect significant excesses above the expected photospheric emission at 21 \micron\ for two of our targets, WD 2149+021 and WD 2105-820. WD 2105-280 has a candidate planetary companion at a projected separation of 34 au and an infrared excess--if both candidates are confirmed, it would represent the first WD multi-planet system. We investigate whether these excesses could be caused by very low luminosity warm dust disks or planets. While both are likely, we argue that the most likely explanation for the excesses is that they are the thermal emission from jovian-mass planets in orbits with semi-major axes $<$10 au, using a combination of observational constraints. If most of the candidate planets presented here are confirmed, it would suggest that metal polluted white dwarfs are frequently orbited by at least one giant planet.
\end{abstract}

\keywords{White dwarf stars(1799), Exoplanets(498), Circumstellar disks(235), James Webb Space Telescope(2291)}


\section{Introduction} \label{sec:intro}
The ultimate fate of our Solar System is one of the fundamental questions asked by humanity. While there is no path for us to directly observe the death of our own Solar System billions of years in the future, we can nevertheless turn to astronomical observations to create a basic picture of how things might proceed both for terrestrial planets, giant planets, and the associated minor bodies and moons that fill out the known Solar System. 

Studying the evolution of planetary systems at the extremes of planet survivability is a useful tool for gaining more insight into the architecture and chemistry of planetary building blocks around main sequence stars. Just as the careful pathologist can study a human body to determine a person's life and habits, so too can astronomers sift through the wreckage of past planetary systems to learn how they are formed. It turns out that careful observation of the end states to main sequence evolution, white dwarfs (WDs), provide many helpful and sensitive avenues to this pursuit. 

Even before WDs were recognized as the corpses of solar-type stars, evidence of atmospheric pollution had been identified \citep{vanMaanen}. The first spectroscopy of WDs occasionally showed Ca absorption features, earning them the designation of DAF or DAwk \citep{eggen65}. These discoveries were initially attributed to accretion from the interstellar medium (ISM) \citep{dupuis92} or radiative levitation in hotter WDs. The advent of visible echelle spectroscopy with large telescopes \citep[e.g.][]{zuckerman03} and UV spectroscopy \citep[e.g.][]{koester14} with the Hubble Space Telescope revealed a significant occurrence of metal pollution around WDs, with over 25\% of WDs showing some level of pollution and that by and large the abundance patterns of the pollution are consistent with rocky bodies as opposed to the ISM \citep[e.g.]{xu14}.

The launch of the James Webb Space Telescope allows a new window into the fate of planetary systems. The MIRI Imager \citep{bouchet.2015, dicken.2024} is sensitive enough to detect cold giant planet companions to WDs \citep{limbach22, mullally24,limbach25} and small amounts of dust \citep{poulsen24}. Giant planets are detectable via both infrared excess and resolved companions, while dust is observed as an infrared excess, since WD dust disks are very compact and reside close to the Roche tidal disruption radius of the host star ($R\sim$1~\Rsun).

As part of an exploratory JWST Cycle 1 Program \citep[PID 1911 PI: S.Mullally][]{mullally.2021jwstprop}, we searched four nearby metal-polluted WDs with no known IR excess for evidence of planetary companions or debris. Despite a small sample size, three of the four WDs show interesting evidence for planets or dust, suggesting that further observations with JWST could reveal a wealth of information about planetary systems that are past their prime.

In \S \ref{sec:sample} we discuss the four WD targets, while in \S \ref{sec:obs} we detail the observational program and report precise mid-IR photometry and imaging of these WDs beyond 12~\micron\ for the first time. In \S \ref{sec:resolved}, we detail direct imaging searches for resolved substellar companions to the four WDs, placing stringent limits to the presence of giant planets in orbits beyond $\sim$10~au. In \S \ref{sec:unresolved} we report excesses around two of the four WDs and determine the most likely interpretation as two candidate unresolved planets in orbit around WD~2149+021 and WD~2105-820 in \S \ref{sec:excessinterp}. Finally we discuss our results in \S \ref{sec:discussion}.

\section{The WD Sample} \label{sec:sample}

The four WDs were selected to contain evidence of metal pollution from at least one atomic species, be nearby to the Earth, and have a total system age less than that of the Earth to ensure a sensitivity to giant planets with the least amount of observing time.  We list details of the targets in order of decreasing distance from the Earth, and report our derived fundamental parameters in \S\ref{sec:photparam}. A brief history of WD~2149+021 was already presented in \citet{poulsen24}, so we skip that object.

\subsection{WD~2105-820}
WD~2105-820 was reported as a WD in \citet{luyten49}. It was identified early on as a potentially good target for directly imaging giant planets \citep{burleigh02} and was soon found to have a detectable Ca line \citep{koester05}, indicative of active rocky material accretion. It was discovered to have a $\sim$9kG magnetic field \citep{landstreet12}, and further measurements imply that it is consistent with a simple dipole with a magnetic axis that is roughly aligned with the rotation axis \citep{bagnulo18}. Interestingly, there is evidence that such a magnetic field can suppress convection in the upper atmosphere of this WD, which might have implications for the inferred accretion rate of metals and does slightly modify the inferred fundamental parameters of this target \citep{gentilefusillo18}. \citet{mullally24} reported a candidate planet between 1-2~\Mjup\ around this WD at a projected separation of 34~au.

\subsection{WD~1620-391}
WD~1620-391 was first discovered to be a DA WD via photographic plate spectra under the name CD-38$\arcdeg$10980 \citep{stephenson_sanduleak67}. It was recognized as a possible common proper motion companion to HR 6094, which was later confirmed
. HR 6094 is also host to a Jupiter analog exoplanet companion in a 1.32~au orbit \citep{mayor03}. It was the third WD to have an estimate of its gravitational redshift \citep{wegner78}, and is constrained to have a magnetic field of $<$7.8kG \citep{kawka07}.
At a distance of 12.911$\pm${0.007}~pc, it is one of the closer hot WDs, making it amenable to studying both its EUV and X-Ray emission \citep[e.g.][]{marsh97}. The local ISM G cloud is in this direction of the sky and has a projected heliocentric radial velocity of $-25$ km s$^{-1}$, well separated from the observed heliocentric radial velocity of the WD (v=44.0$\pm$0.6~km~s$^{-1}$)\citep{malamut14,napiwotzki20}. WD~1620-391 has also been used as a target for previous direct imaging searches for planets in the NIR and mid-IR with HST and Spitzer \citep{debes05,dodo2}, with no compelling companions or excesses detected \citep{debes07}.

WD~1620-391's status as a DAZ has historically been muddied in the literature. Optical echelle spectroscopy shows no obvious Ca H or K lines, nor are any trace metal lines obviously detected. UV observations from IUE (International Ultraviolet Explorer) in particular seemed to show only circumstellar velocities for transitions of Si due to a difference between the observed line centers and that of hydrogen lines in the visible \citep{holberg95}. FUSE (Far Ultraviolet Spectroscopic Explorer) observations of WD~1620-391 showed the presence of Si III lines near 1100~\AA\ that were consistent with a photospheric origin; an assumed log Si/H abundance of $-7.69$ in the atmosphere predicted consistent equivalent widths in Si for both the FUSE and the IUE spectra\citep{wolff01}. These authors could not reconcile the possible velocity differences to that expected in the photosphere. Given that the IUE data measured the radial velocity of the ISM components to be $-30$~km s$^{-1}$, the most likely explanation is that there was a +5~km~s$^{-1}$ velocity offset in the IUE data. When that is accounted for, the average velocity of the reported circumstellar components is 44~km~s$^{-1}$; in line with the latest photospheric RVs derived from optical echelle spectroscopy \citep{napiwotzki20}. We thus consider WD~1620-391 a bonafide DAZ. 

To definitively confirm this, we obtained the Hubble Advanced Spectral Products (HASP) Space Telescope Imaging Spectrograph (STIS) UV echelle spectra of WD~1620-391, which includes data from HST program ID 14076 \citep[PI: Gaensicke; E140M, E230M][]{debes24}. We focused our attention on the species reported in the IUE data and verified whether they were detected, and measured their radial velocities. We obtained a Gaussian fit to each of 20 candidate photospheric lines and 15 ISM lines, obtained the best fit line center, and then calculated a radial velocity from those fits. The mean value and the standard deviation in measured velocities from the STIS data for the ISM components is -26.6$\pm$1.8~km~s$^{-1}$. All of the non-ISM lines reported in \citet{holberg95} were also strongly detected in the STIS observations and the mean of their velocities was 43.52$\pm$1.3~km~s$^{-1}$. Both values are consistent, within the uncertainties, to previous determinations of the ISM velocity in this line of sight and to the radial velocity of WD~1620-391 as determined in the visible, confirming that active rocky accretion is occurring. 

\subsection{WD~1202-232}

WD~1202-232 was first identified as a DA as part of the Edinburgh-Cape Survey \citep{kilkenny97}, and then was noted to have Ca metal line absorption shortly after its discovery \citep{zuckerman03,koester05}. It has been searched for magnetic fields but no significant field was detected \citep{jordon07,bagnulo18}. Low resolution spectroscopy from the UV through the near-IR has been obtained for this target with HST as part of the CALSPEC series of high quality spectrophotometric stars \citep{bohlin}, and a candidate planetary companion between 1-7~\Mjup\ was reported in \citet{mullally24} at a projected separation of 11~au.

\section{Observations} \label{sec:obs}
JWST program ID 1911 targeted the WDs with the mid-infrared instrument (MIRI) in full readout, imaging mode with four different broadband filters: F560W, F770W, F1500W, F2100W.  The observations' exposure times were designed to make it possible to directly image planets with masses greater than or equal to 1 Jupiter mass orbiting the WD with the 15~\micron\ filter. The 21~\micron\ filter exposure was set to ensure a signal to noise of 55 in order to detect unresolved planets larger than one Jupiter mass using infrared excess. The short exposures of the 5 and 7\,\micron\ filters were intended to reject false-positive detections of non-planetary resolved companions, and to measure the white dwarf photospheric brightness against which flux excesses from unresolved companions can be detected. For each filter a 4, 8 or 12-point dither in the cycling pattern were used: a 4-pt pattern for the short F560W and F770W observations, a 12-pt pattern for the deep F1500W observations of WD~2105-820 and WD~2149+021 and a 8-pt pattern for all the rest.

 \begin{figure}
    \plotone{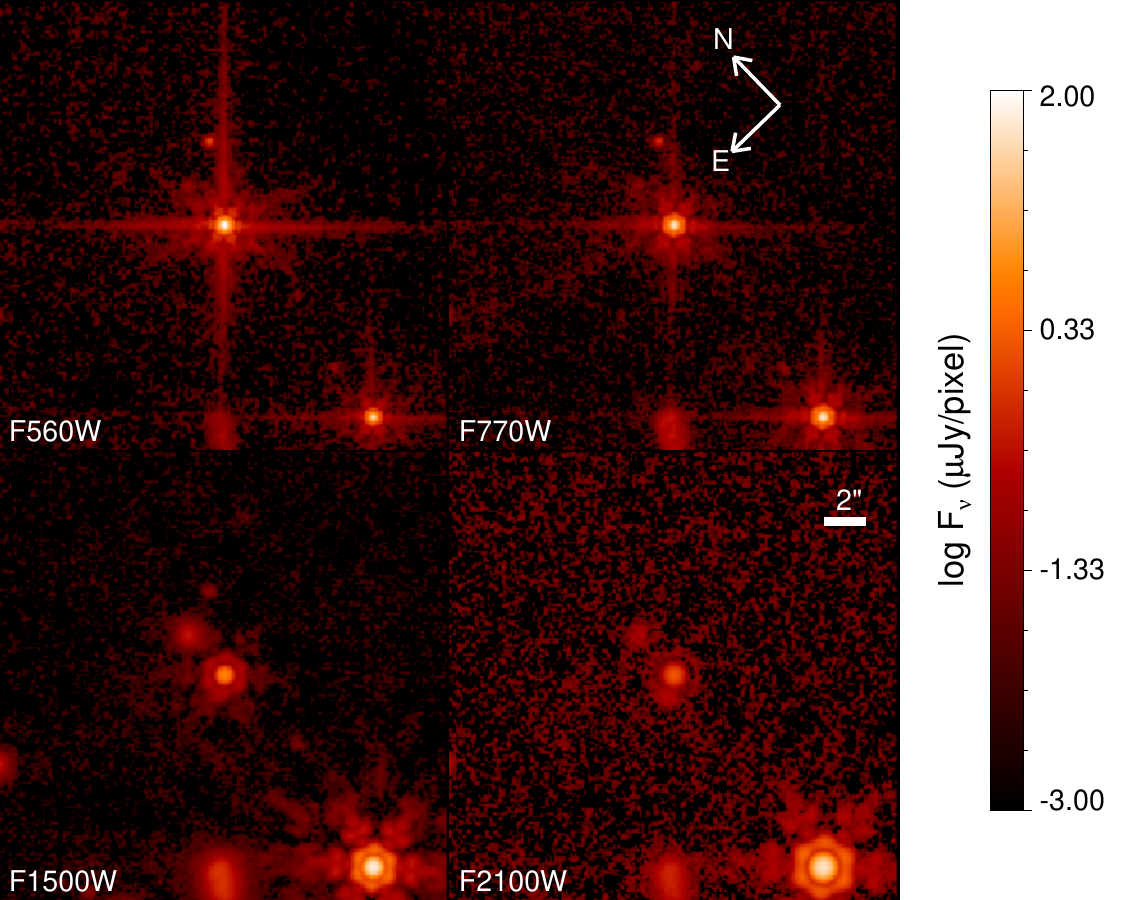}
    \caption{\label{fig:im2} The four MIRI images of WD~1202-232. Each field represents 22\arcsec$\times$22\arcsec\ with the WD at the center.}
\end{figure}

The images were processed using build 11.1 (data processing software version 2022\_5b and calibration software version of 1.16.0) of the JWST Calibration pipeline starting from the uncal files, using a CRDS context from jwst\_1293.pmap and CRDS version of 11.18.4. Each set of images was processed through stage one (calwebb\_detector1) and two (calwebb\_image2) of the imaging pipeline using mostly default parameters. The only non-default values set were in the jump step of the calwebb\_detector1 pipeline. For that step, we set the rejection threshold to 5~$\sigma$ rather than the default of 4, and turned off the cosmic ray shower finding code, using only the standard jump detection algorithms. After the stage two pipeline was run, a median background image was created from the stacked \*\_cal.fits images for each filter and subtracted from each cal file. Lastly, the level three imaging pipeline (calwebb\_image3) was run on the background subtracted files and combined with the resample kernel set to `square', the resample weight\_type set to ‘exptime’,  and the outlier detection `scale' values being set to (1.0 and 0.8), which is twice the default values, but using the standard parameter reference files for all other parameter settings. These background subtracted cal and i2d files were used for all subsequent analyses.

\begin{figure}
    \plotone{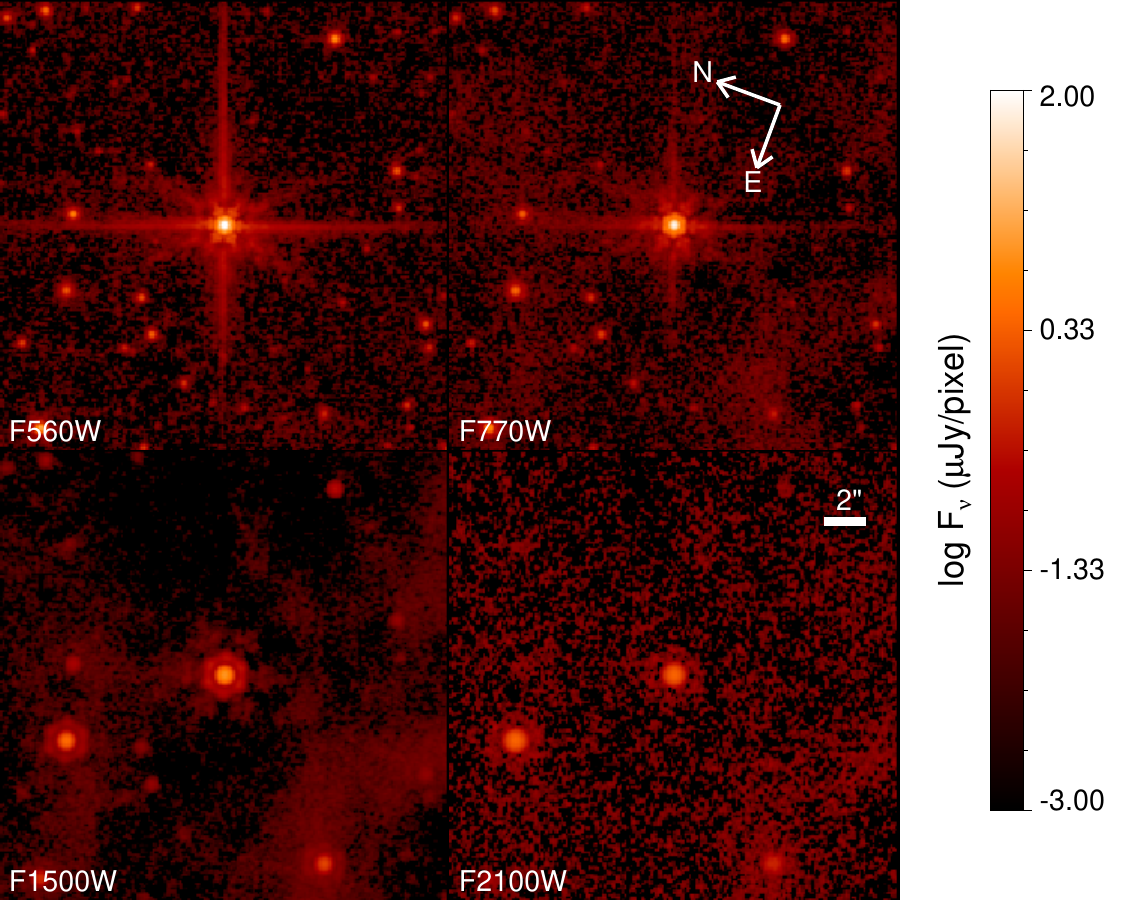}
   \caption{\label{fig:im1} The same as Figure \ref{fig:im2} but for WD~1620-391.}
\end{figure}
 
Most of the updates for the build 11.1 pipeline were minor bug fixes and code cleanup like ensuring consistency between the SCI, ERR and DQ extensions of the data to make sure that a pixel flagged as DO\_NOT\_USE in the DQ extension would be flagged as a bad pixel and not used in the SCI and ERR arrays. These updates would not have had a large effect on the data from previous versions of the pipeline, though there may have been a few more Not a Number valued pixels in the calibrated files before combination in the (calwebb\_image3) pipeline.

\begin{figure}
    \plotone{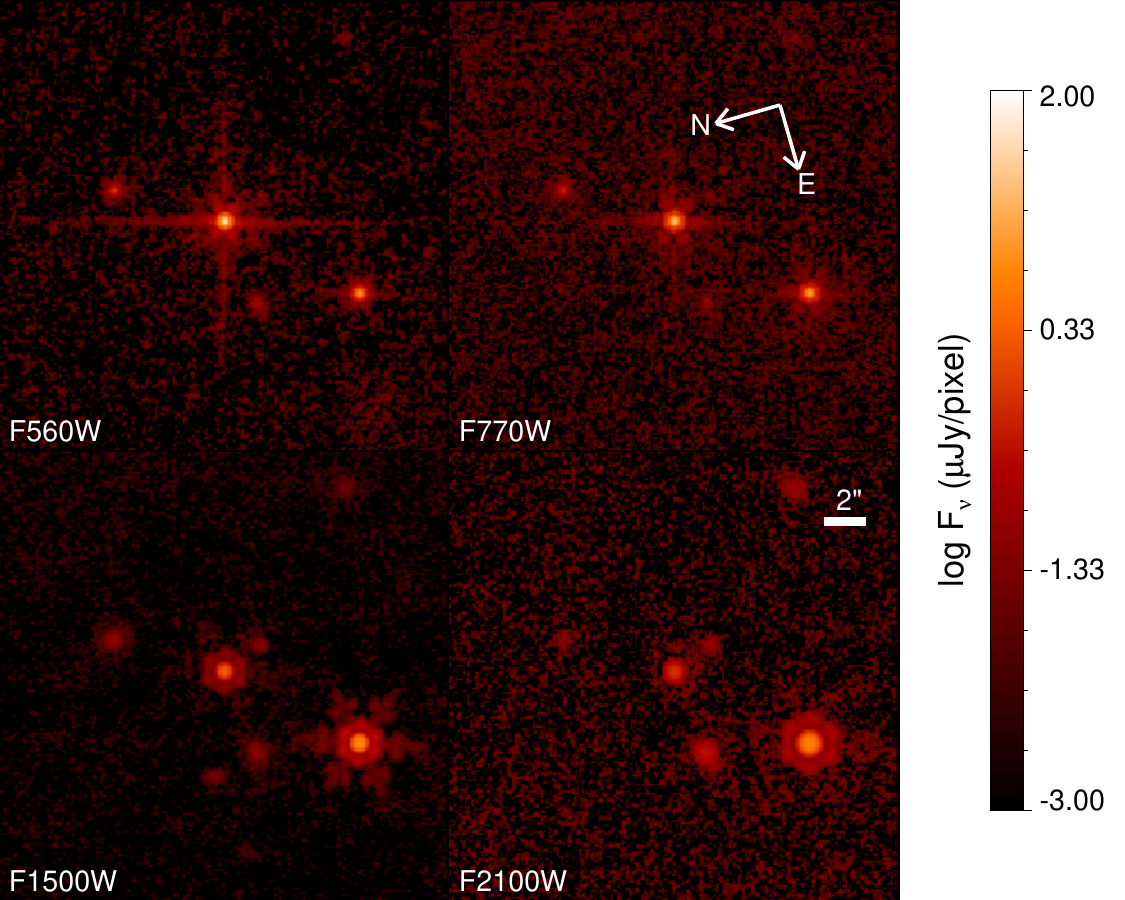}
    \caption{\label{fig:im3} The same as Figure \ref{fig:im2} but for WD~2105-820.}
\end{figure}

\begin{figure}
    \plotone{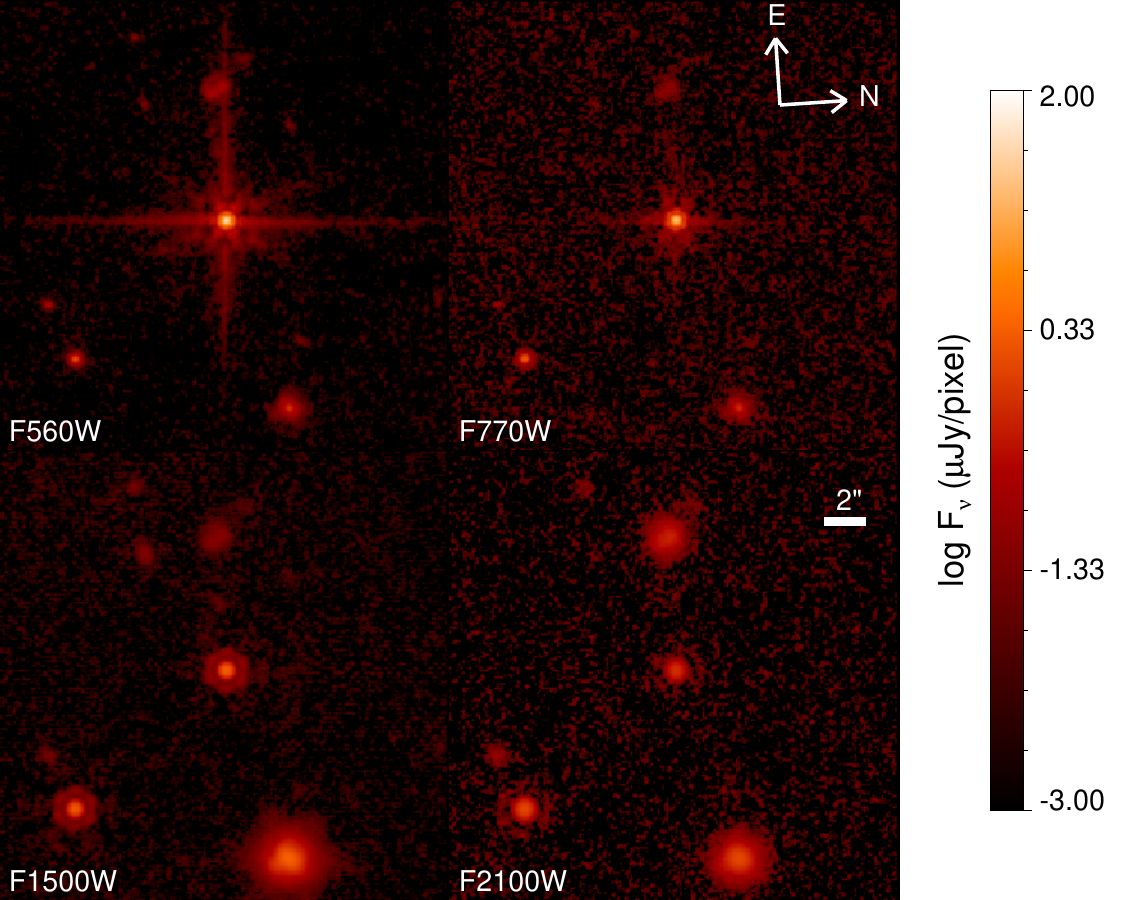}
    \caption{\label{fig:im4} The same as Figure \ref{fig:im2} but for WD~2149+021.}

\end{figure}

Figures \ref{fig:im2} to \ref{fig:im4} show 22\arcsec$\times$22\arcsec\ fields surrounding each WD target in the four filters. With the exception of WD~1620-391, the fields contain a mix of galaxies of varying morphology and a smattering of point sources. Since WD~1620-391 is close to the galactic plane ($b$=+7.3 degrees), some spatially variant galactic cirrus is most noticeable at 7 and 15~\micron, complicating the fidelity of the background subtraction and decreasing our sensitivity to faint point sources in localized regions. 

\subsection{White Dwarf Photometry} \label{sec:photparam}
Postage stamp images from the dither-combined images were created of each WD to ensure that aperture photometry could be performed to match the procedure obtained for the MIRI absolute flux calibration \citep{gordon24}. The default aperture sizes and background annuli for each filter that corresponded to 70\% enclosed energy was derived using the aperture correction file\ jwst\_miri\_apcorr\_0014.fits located at the JWST CRDS website\footnote{https://jwst-crds.stsci.edu/}. We made use of the IDL routine {\em aper.pro} part of the IDL Astronomy Users Library\footnote{https://github.com/wlandsman/IDLAstro}. Based on this approach, our photometry should be accurate to $\sim$2\%, with this uncertainty dominated by any systematic uncertainties between our predicted photospheres and the observed photometry. In \S \ref{sec:excess}, we look more deeply into what our likely systematic uncertainties might be.

In addition to the JWST photometry we compiled visible and Near-IR photometry for each WD, similar to the procedure for WD~2149+021 detailed in \citet{poulsen24}. For WD~2105-820, we made use of GALEX \citep{bianchi2017}, Sloan $g/r/i/z$,APASS DR9 B/V, Gaia DR3 G/G$_{BP}$/G$_{RP}$, 2MASS J/H/K, ALLWISE W1/W2/W3, and Spitzer IRAC1, IRAC2, IRAC3, IRAC4 SEIP fluxes. We obtained a best fit T$_{eff}$, $\log{g}$ based on the Gaia parallax and a chi-square minimization against the publicly-available cooling models provided by P. Bergeron \citep{holberg06,bedard20} and report the results in Table \ref{tab:wdparams}. WD~1202-232 also has an empirical CALSPEC spectrum, and as mentioned previously, these spectra have absolute flux accuracies that approach 1\%. To that end, we convolved the CALSPEC spectrum, which spans the FUV to the Near-IR, into the GALEX and Sloan $g/r/i/z$ filters using the Spanish Virtual Observatory SPECPHOT tool\footnote{http://svo2.cab.inta-csic.es/theory/specphot/}. WD~1620-391 possesses a high SNR STIS/E140M+E230M spectrum that we used to calculate synthetic GALEX photometry, as it is too bright for accurate GALEX fluxes. Some bands were ignored due to possible saturation, contamination from atmospheric extinction such as the Sloan $u$, possible contamination from background sources such as W3, or in the case of WD~2105-820, GALEX fluxes were ignored to better approximate the magnetic radiative model required, which produces different photospheric fluxes in the UV compared to a standard non-magnetic convective model \citep{gentilefusillo18}.

\begin{deluxetable*}{lcccc}
\label{tab:wdphot}
\caption{Photometry of the target WDs} 
\tablehead{
\colhead{Filter} & \colhead{WD2149 Flux} & \colhead{WD2105 Flux} & \colhead{WD1620 Flux} & \colhead{WD1202 Flux} \\
\colhead{} & \colhead{$\mu$Jy} & \colhead{$\mu$Jy} & \colhead{$\mu$Jy} & \colhead{$\mu$Jy}
}
\startdata
GALEX FUV &  50000$\pm$   500 & & 524725$\pm$ 15741 &    184$\pm$     1 \\
GALEX NUV &  39799$\pm$   397  &   6309$\pm$   174 & 341963$\pm$ 10258 &   8408$\pm$    84 \\
g$^\prime$ &  32342$\pm$   646 & & 26507$\pm$   530 \\
r$^\prime$ &  23816$\pm$   476 & & 27051$\pm$   541 \\
i$^\prime$ &  18241$\pm$   364 & & 25528$\pm$   510 \\
z$^\prime$ &  13836$\pm$   276 & & 22731$\pm$   454 \\
U &  & & 216796$\pm$  3993 &  \\
B &  33025$\pm$  1247 &  11440$\pm$   579 & 181654$\pm$  3346 &  24257$\pm$   670 \\
V &  28250$\pm$  1717 &  13274$\pm$   501 & 143689$\pm$  2646 &  28094$\pm$   776 \\
R &       & & 107916$\pm$  1987 & \\
I &       & &  74661$\pm$  1375 &  \\
G$_{\mathrm{BP}}$ &  28258$\pm$   565 & 11769$\pm$   235 & 152455$\pm$  3049  &  25092$\pm$   501 \\
G &  25157$\pm$   503 & 11712$\pm$   234 & 129139$\pm$  2582 &  26053$\pm$   521 \\
G$_{\mathrm{RP}}$ &  17269$\pm$   345 & 10122$\pm$   202 &  82503$\pm$  1650  &  24551$\pm$   491 \\
J &   8365$\pm$   184 &  6475$\pm$   155 &  37298$\pm$   721 &  17477$\pm$   386 \\
H &   4964$\pm$   169 & 4264$\pm$   129 &  21237$\pm$   449 &  12311$\pm$   306 \\
K$_{s}$ &   2918$\pm$    99 &  2582$\pm$    92 &  13085$\pm$   265 &   7726$\pm$   185 \\
W1 &   1369$\pm$    31 & 1278$\pm$    29 &   5596$\pm$   118 &   3660$\pm$    84 \\
W2 &    764$\pm$    22 & 703$\pm$    19 &   2793$\pm$    59 &   1985$\pm$    43 \\
W3 &      & 236$\pm$   105 & &    643$\pm$   138 \\
IRAC1 &     & &   4975$\pm$   149 &  \\
IRAC2 &    786$\pm$    23 & 727$\pm$    21 &   3119$\pm$    93 &  2033$\pm$    60 \\
IRAC3 &    & &   1974$\pm$    59 &  \\
IRAC4 &    272$\pm$     8 &  256$\pm$     7 &   1076$\pm$    32 &    700$\pm$    21 \\
F560W & 508 $\pm$ 10 & 463 $\pm$ 9 & 2030 $\pm$ 41 & 1304 $\pm$ 26\\
F770W & 282 $\pm$ 6 & 262 $\pm$ 5 & 1112$\pm$ 22 & 726 $\pm$ 14 \\
F1500W & 73 $\pm$ 2 & 71$\pm$ 1 & 277 $\pm$ 6 & 183 $\pm$ 4\\
F2100W & 43 $\pm$ 1 & 42$\pm$ 1 & 145 $\pm$ 3 & 103 $\pm$2 \\
\enddata
\end{deluxetable*}

\begin{deluxetable*}{lcccc}
\label{tab:wdparams}
\caption{Fundamental parameters of the target WDs} 
\tablehead{
\colhead{} & \colhead{WD2149} & \colhead{WD2105} & \colhead{WD1620} & \colhead{WD1202}
}
\startdata
d(pc) & 22.56$\pm$0.02 & 16.179$\pm$0.004 & 12.91$\pm$0.01 & 10.427$\pm$0.002 \\
T$_{\mathrm eff}$ (K) & 17840$\pm$160 & 9890~K\tablenotemark{a} & 24700$\pm$600 & 8720$\pm$120 \\
$\log{g}$ & 8.01$\pm$0.01 & 8.20$\pm$0.03\tablenotemark{a} & 8.04$\pm$0.03 & 8.00$\pm$0.04 \\
M (\Msun) & 0.62$\pm$0.01 & 0.7$\pm$0.06 & 0.64$\pm$0.01 &  0.6$\pm$0.03 \\
t$_{\mathrm cool}$ (Gyr) & 0.114$\pm$0.003 & 0.83$\pm$0.1 & 0.026$\pm$0.004 & 0.90$\pm$0.07 \\
t$_{\mathrm MS}$ (Gyr) & 2.9$^{+3.7}_{-1.3}$ & 1.6$^{+0.8}_{-0.24}$ & 2.18$^{+3.04}_{-0.84}$  & 5.24$^{+5.04}_{-2.50}$  \\
\enddata
\tablenotetext{a}{Based on the parameters determined in \citet{gentilefusillo18}}
\end{deluxetable*}

\subsection{Procedure for predicting JWST MIRI Fluxes} 
The publicly available Bergeron cooling models do not have predictions for any of the MIRI bands, and thus we took the best fit WD parameters and used them to generate high resolution photospheric models specific to each WD that extended to 30\micron. We first scaled the photospheric model by the best fit radius and distance, and then further scaled these spectra to existing CALSPEC or STIS UV spectra. WD~2105-820 was not corrected at this step. In general this second step changed things by less than 1\%.

Then, taking the publicly available MIRI filter transmission profiles from the Spanish Virtual Observatory Filter Service \citep{rodrigo2024}  we convolved the models with each profile to create an accurate prediction of the MIRI flux in each band \citep{holberg06}.

\begin{figure}
    \plotone{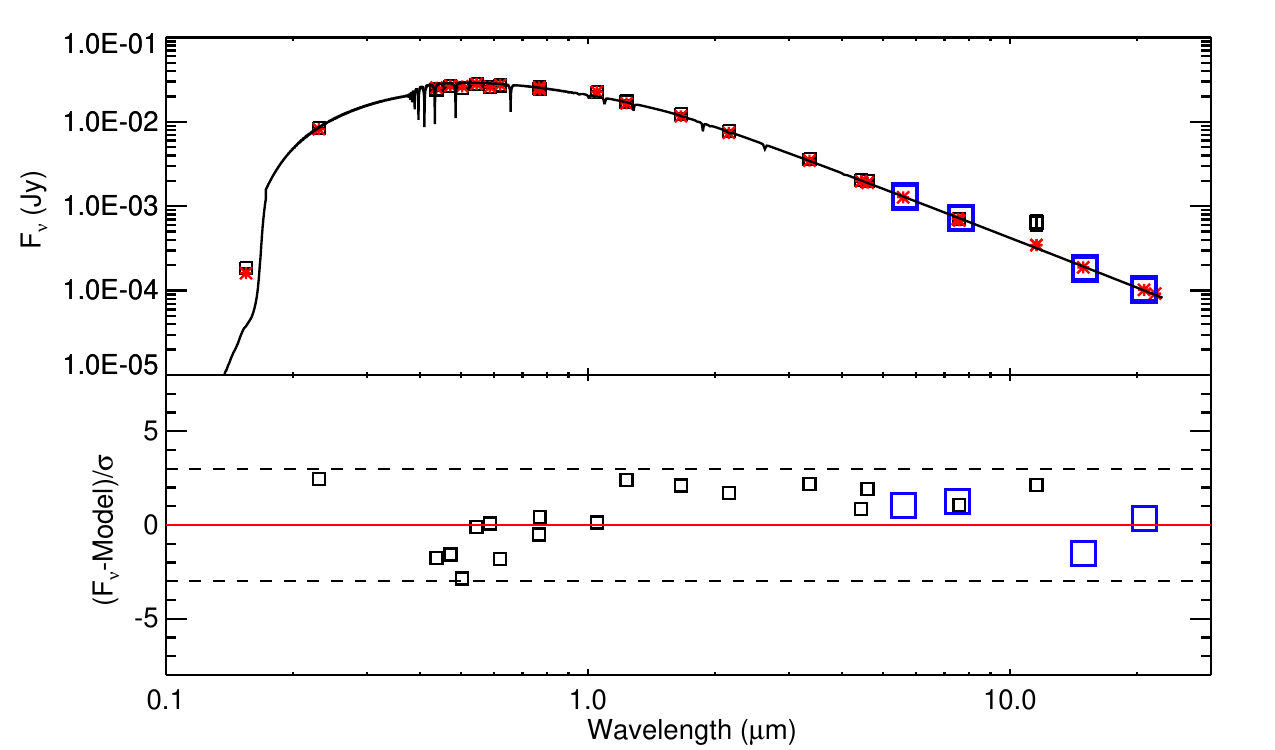}
   \caption{\label{fig:sed1202} (top panel) Spectral energy distribution of WD~1202-232 and comparison to its predicted photospheric emission. The black line represents the model photosphere for the WD, the red asterisks are the predicted model fluxes in a given filter, and the black squares are archival photometry for the WD (See Table \ref{tab:wdphot}). The blue squares represent the observed JWST fluxes in F560W, F770W, F1500W, and F2100W. (bottom panel) Residual photometry relative to the model photosphere, weighted by the photometric uncertainties. The black squares are archival photometry, while the blue squares represent the JWST photometry.}
  
\end{figure}

\begin{figure}
    \plotone{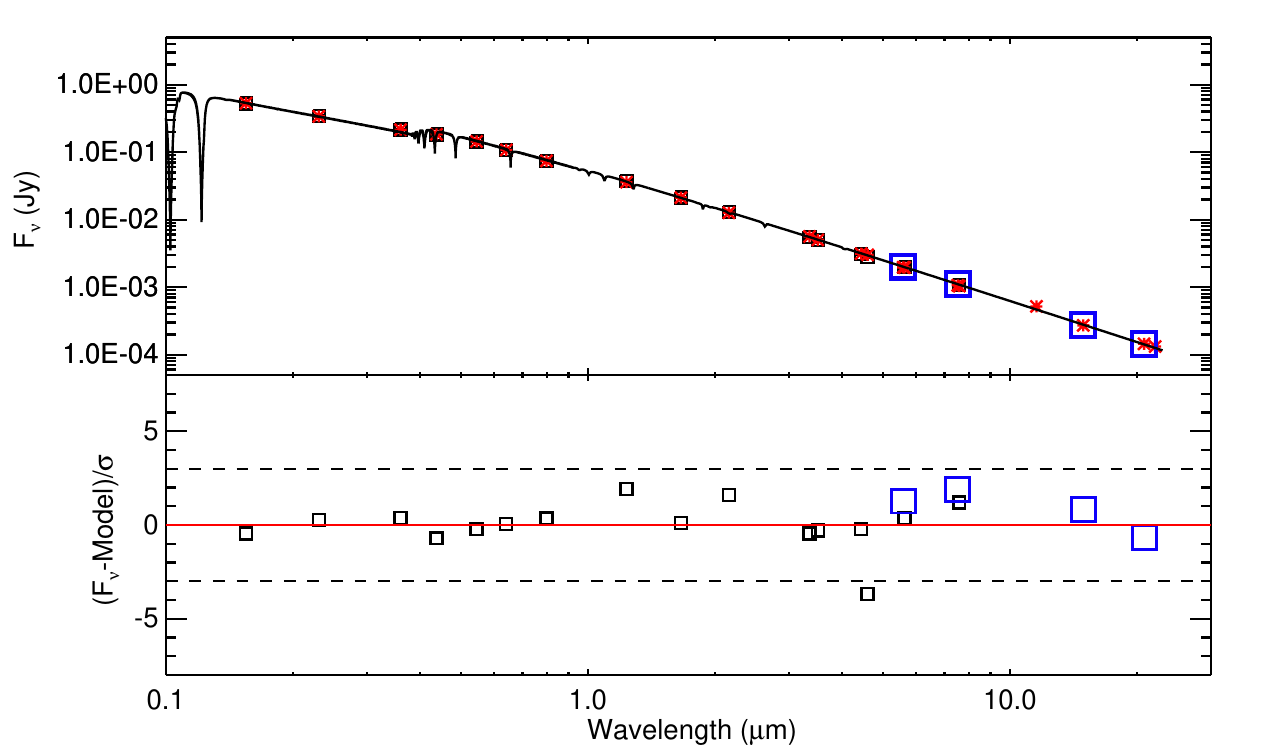}
    \caption{\label{fig:sed1620} Same as Figure \ref{fig:sed1202}, but for WD~1620-391.}
\end{figure}

Figures \ref{fig:sed1202} to \ref{fig:sed2149} show the resulting spectral energy distributions (SEDs) for each of the four WDs compared to both the Bergeron photometric predictions (red asterisks), the extended photospheric spectral models (black lines), the observed archival photometry (black squares), and the new MIRI photometry (blue large squares). A comparison between observed and predicted flux in each filter bandpass is also shown in the lower panel of each figure with $\pm$3-$\sigma$ boundaries denoted as dashed lines. 

\subsection{Photometry match to the models} \label{sec:excess}
\begin{figure}
    \plotone{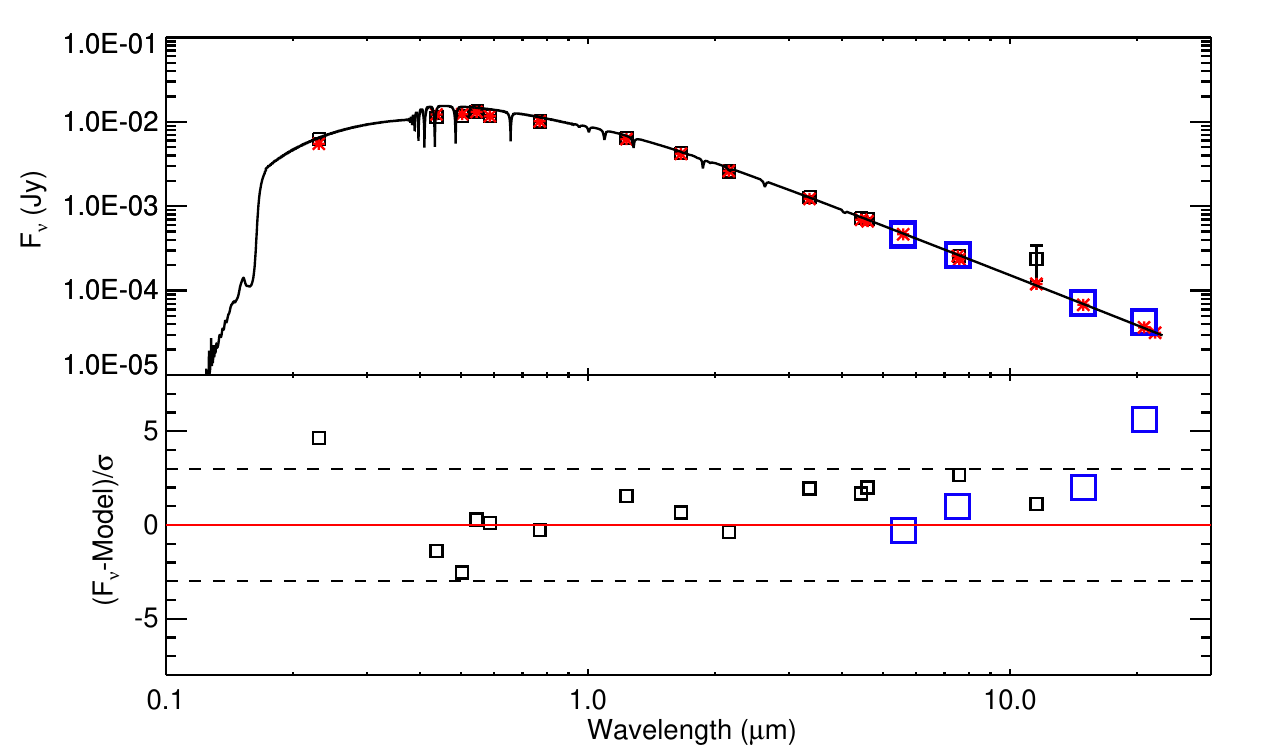}
    \caption{\label{fig:sed2105} Same as Figure \ref{fig:sed1202}, but for WD~2105-820. Note the significant 5.6~$\sigma$ excess at 21~\micron. Additionally, there is a tenatative excess at 2.0~$\sigma$ at 15~\micron.}
\end{figure}

\begin{figure}
    \plotone{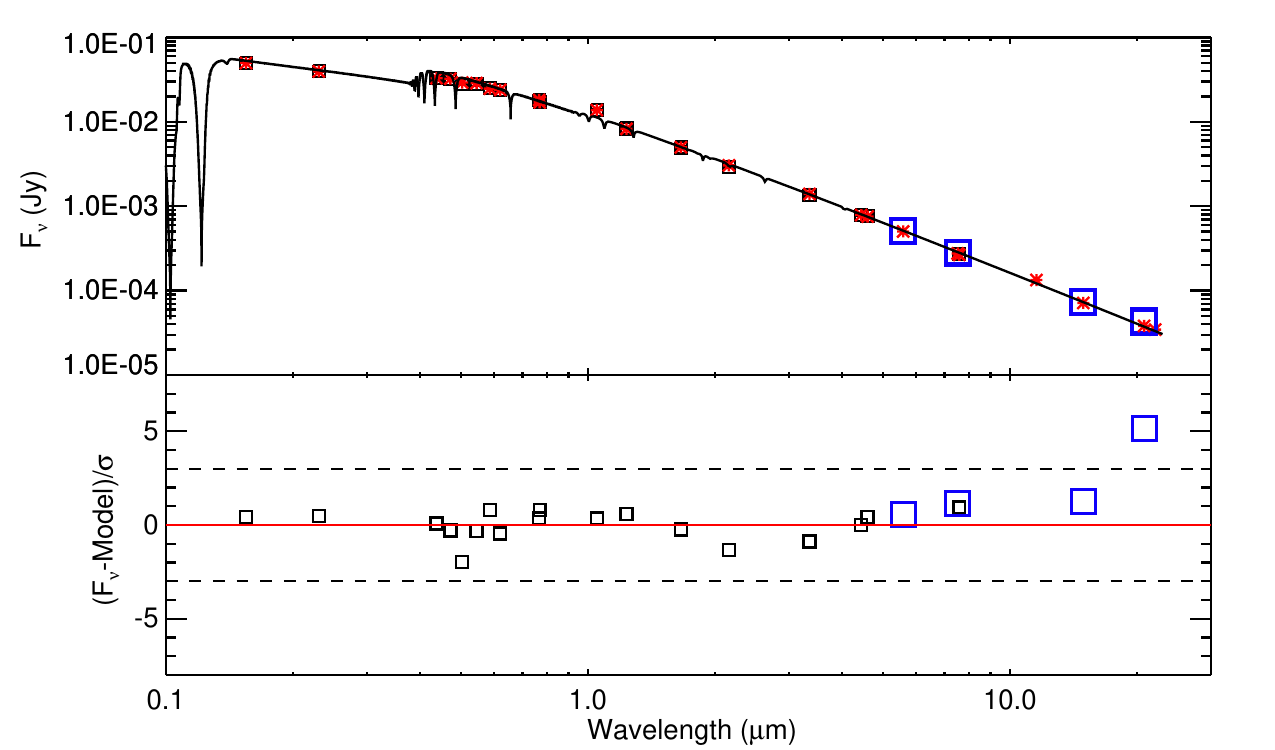}
    \caption{\label{fig:sed2149} Same as Figure \ref{fig:sed1202}, but for WD~2149+021. Note the significant 5.1~$\sigma$ excess at 21~\micron.}
\end{figure}

When we compare the WD model photospheres to the observations we find that WD~1620-391 and WD~1202-232's photometry match to within 3-$\sigma$ across all wavebands, including the JWST MIRI filters, with the assumption of 2\% systematic uncertainties combined with the existing photometric uncertainties. For WD~2105-820 and WD~2149+021, we find a similar match across all wavebands except for their F2100W photometry, which deviate from the models significantly by 5.6 and 5.1~$\sigma$ respectively. Additionally, both targets' F1500W photometry is elevated at 2.0 and 1.2~$\sigma$, respectively. See Figures~\ref{fig:sed2105} and \ref{fig:sed2149}.

To estimate our systematic photometric uncertainties, we try to replicate the procedures adopted for the absolute flux calibration of MIRI \citep{gordon2022}, we also downloaded archival data and CALSPEC models of JWST WD calibration standards \citep{gordon2022}, G191-B2b (F560W, F770W, F1500W), LHS~749b (F560W, F770W), GD~71 (F560W, F770W), GD~153 (F560W, F770W), and WD~1057+719 (F560W, F770W). To supplement F1500W, we included three F1500W images of the brightest DAs from the MEAD survey and created our own models for them: WD~0839-327,  WD~0752-676, and WD~0644+375 (Program 3964, PI:Poulsen).
Since none of the WD standards had F2100W images, we also inspected three individual full frame F2100W observations of BD+60-1735, as well as a SUB64 F2100W observation of 10 Lac, $\mu$ Col, and $\lambda$ Lep, all part of the hot stars sample from \citet{gordon2022}. We also included an F2100W observation of WD~0644+375 from the MEOW survey (Program 4403, PI: Limbach).  This ensured we had between 7 and 10 independent measures of MIRI flux in all four filters compared to at least 7 different expected model photospheres. In every filter except F2100W our sample was derived from only white dwarfs, with F2100W using a sample with half white dwarfs and half hotter stars.

In the case of LHS 749B, the F560W flux measured for this star was anomalously low compared to the best guess model. This could be due the possibility that for some of the images smaller dither offsets were used, which can increase negative flux residuals at the sky subtraction step of calibration. Additionally, there is evidence that LDS 749B is fainter than expected around 5~\micron\ in NIRSPEC calibration data, indicating the possibility of additional atmospheric opacity that the current best model for the star does not include (C. Proffit, private communication). We did not include the F560W photometry from LDS 749B in our estimates below.

We then performed the same aperture photometry for all of the calibration targets and our targets (with the exception of the excess candidates in F2100W and WD~2105-820 in F1500W) and calculated the mean and RMS of the measurements, which provides the statistical plus systematic uncertainties in an individual measurement of absolute flux for a given target.

\begin{figure}
    \plotone{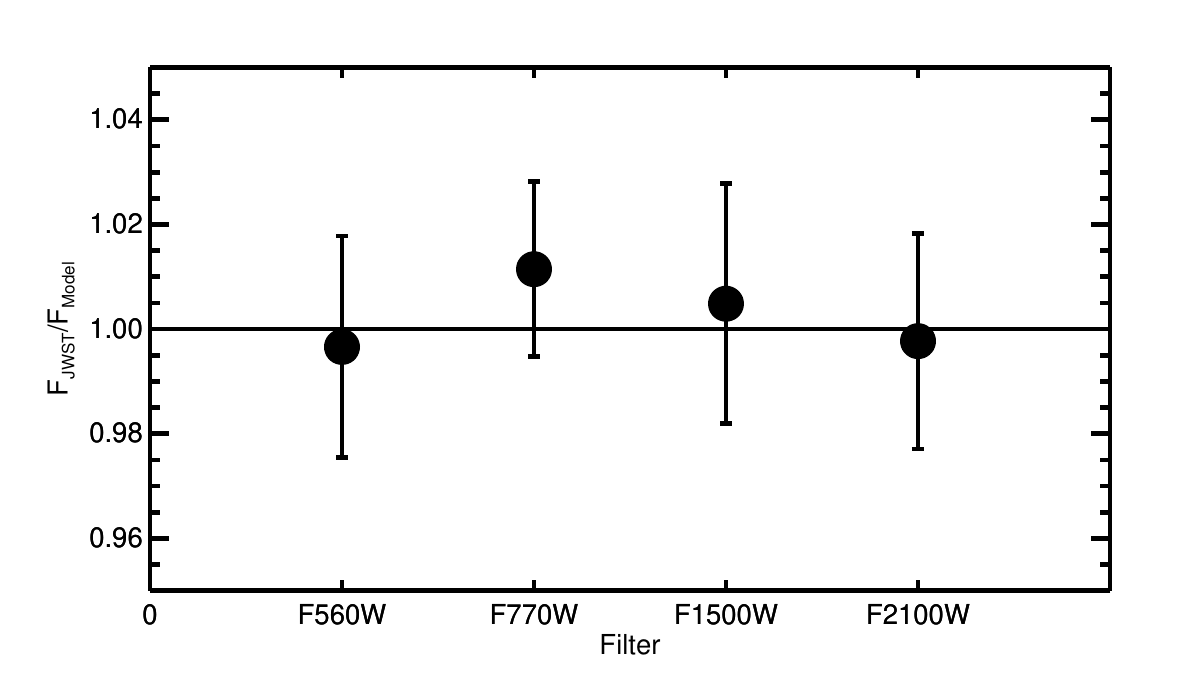}
   \caption{\label{fig:calfactors} Comparison between the average measured flux in a given MIRI imager filter and the expected flux assuming the best-guess models for each calibrator star. Error bars are the calculated rms of each sample of stars, which indicates the systematic uncertainty due to stellar and white dwarf models.}
\end{figure}

Figure \ref{fig:calfactors} shows our resulting independent test of the JWST absolute flux calibration for these MIRI filters. We see an excellent match, in line with an object-to-object RMS of $\sim$2\%. We found that the average ratio of observed photometry to photosphere prediction was 1.00$\pm$0.02 for F560W (8 targets; 8 measurements); 1.01$\pm$0.02 for F770W (7 targets; 7 measurements); 1.00$\pm$0.02 for F1500W (7 targets; 7 measurements); and 1.00$\pm$0.02 for F2100W (7 targets; 10 measurements). This excellent agreement across a factor of $\sim$10 in effective temperature for the WDs suggests that it's highly unlikely that the excesses we measure in the F2100W filter are spurious or due to uncertainties in the absolute flux calibration of JWST or in the models of these well understood WDs.

\begin{figure}
    \plotone{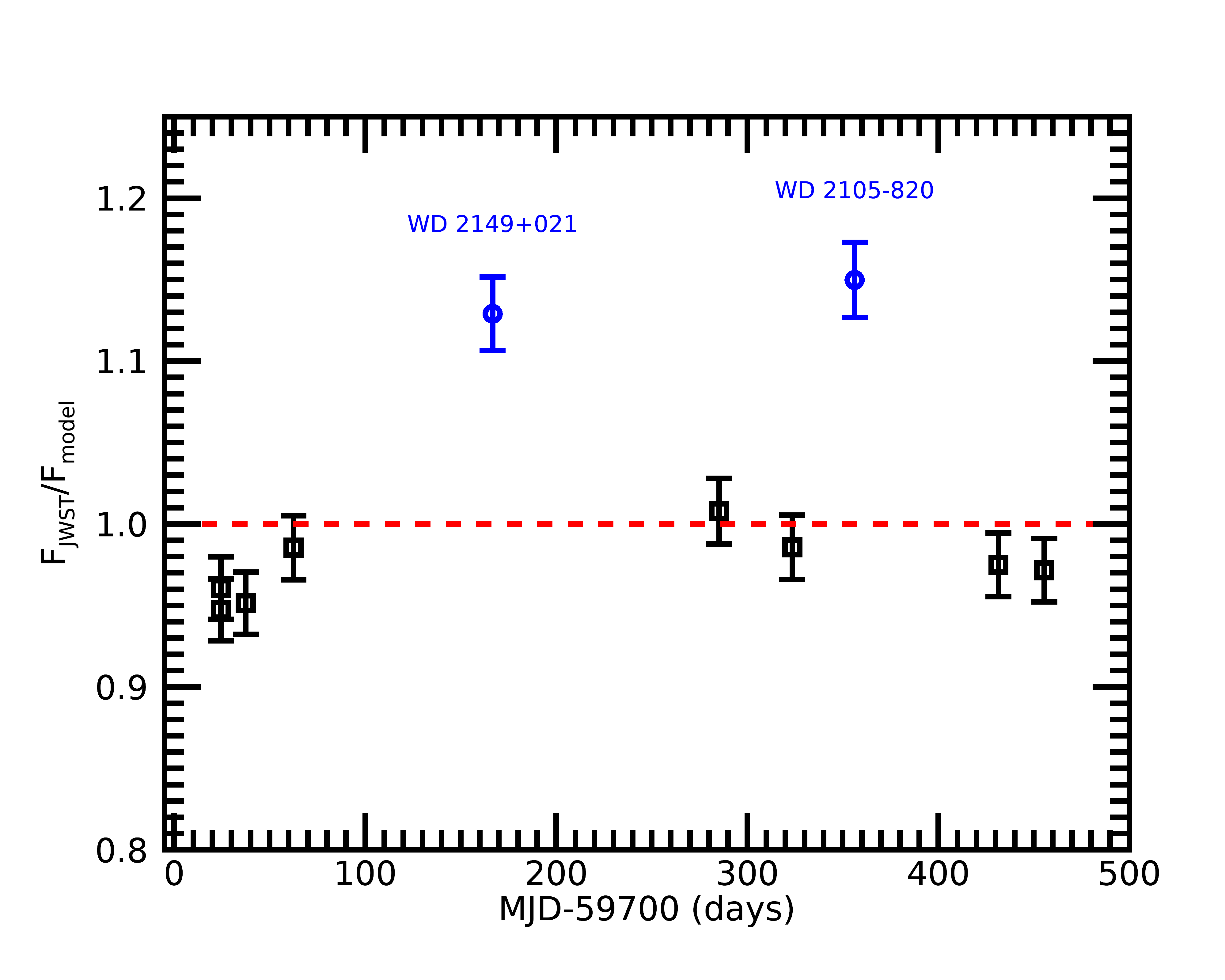}
   \caption{\label{fig:caltime} The ratio of observed flux in F2100W vs. time for calibrators and non-excess WDs (black squares) compared to our excess candidates (blue circles). The calibrator BD+60-1735 was observed starting around MJD 59700 with four epochs, and another two epochs were obtained around 480 days later. The GO 1911 targets were observed between MJD 59850 and 60100. The dashed line is a constant ratio of 1: the median of all observations is 0.985.  The excess candidates are significantly above the constant ratio and there is no strong evidence for any systematic trends due to the sensitivity correction.}
\end{figure}

Another possibility is that the MIRI time-dependent sensitivity was somehow incorrectly applied during the period of time that our WD targets were observed at F2100W, resulting in a subtle systematic offset in the expected flux\footnote{https://jwst-docs.stsci.edu/jwst-mid-infrared-instrument/miri-performance/miri-sensitivity\#gsc.tab=0}. The declining sensitivity was modeled as an exponential decline with time during Cycle 1 and 2. We find the possiblity that an incorrect sensitivity correction is responsible for our observations to be highly unlikely. First, if this were due to a trend in the sensitivity correction, we would expect to see that the excess observed would be correlated with observing epoch. This is not the case--WD~2149+021 was the first WD observed, and WD 2105-820 was the last one. To further investigate this, we measured the photometry of BD+60-1735, which was observed a total of six times within a few months of our WDs in F2100W (See Figure \ref{fig:caltime}). With the exception of the excess WDs, the median ratio of observed to model fluxes was 0.985$\pm$0.013, consistent with no systematic trend over time. Thus we expect the sensitivity correction to add negligible uncertainty to our F2100W photometry.

We note that \citet{poulsen24} did not make a claim for a signficant excess at F2100W for WD~2149+021. This is because for that work they assumed an interpolation between predicted fluxes in the WISE bands rather than computing a full spectral resolution model that could be convolved with the JWST filter transmission curves. We determined that the previous approach overestimates the mid-IR WD continuum by 5-10\%, which we confirmed in tests with G191-B2b at 10 and 15~\micron. This was enough to suppress the long wavelength excess that is present relative to the uncertainties.

Irrespective of the flux calibration factors, relative fluxes bewteen two WDs of similar \Teff\ should be nearly constant in the MIRI filters. We first compare the ratio of WD~2105-820's flux to WD~1202-232's flux. The ratios for F560W, F770W, F1500W, and F2100W are 0.354, 0.361, 0.386, and 0.409. The F2100W ratio is 15\% higher, similar to the excess seen relative to WD~2105-820's model. WD~2149+021 and WD~1620-391 are hotter WDs, their ratio is 0.251, 0.254, 0.265, 0.302 for F560W, F770W, F1500W, and F2100W. WD~2149+021's F2100W ratio is 20\% higher than at shorter wavelengths as well. 


\subsection{PSF Subtraction}

We used the same procedure for PSF subtraction as detailed in \citet{poulsen24} and \citet{mullally24}. Briefly for each WD a 15~\micron\ reference PSF was constructed from an average of WD~1620-391, WD 2149+021, and WD~2105-820 (removing whichever WD was the target) with background sources masked out. The reference PSF was shifted and scaled to match the target WD and subtracted. The images were inspected for significant point sources and all candidates found within a projected physical separation of 100~au had their photometry measured. Anything with a F1500W/F560W ratio $>$1 and a FWHM within 20\% of the expected F1500W FWHM was considered a possible planet candidate. 
 
\subsubsection{Searching for Planets with Kernel Phase Imaging}

Kernel phase imaging (KPI) is an analysis technique that generalizes the concept of aperture masking interferometry to full pupil imaging (regular images), enabling probing for companions at roughly half the traditional diffraction limit \citep{martinache.2010}. The technique performs well when searching for companions at moderate contrasts ($\leq 6-7$\,mag) in the near infrared \citep{pope.2013, kammerer.2019, kammerer.2023} but, here, we apply KPI for the first time in the mid-infrared with MIRI, in the F1500W filter for which exposure times were maximized.

\begin{figure}
    \centering
    \includegraphics[width=1.0\linewidth]{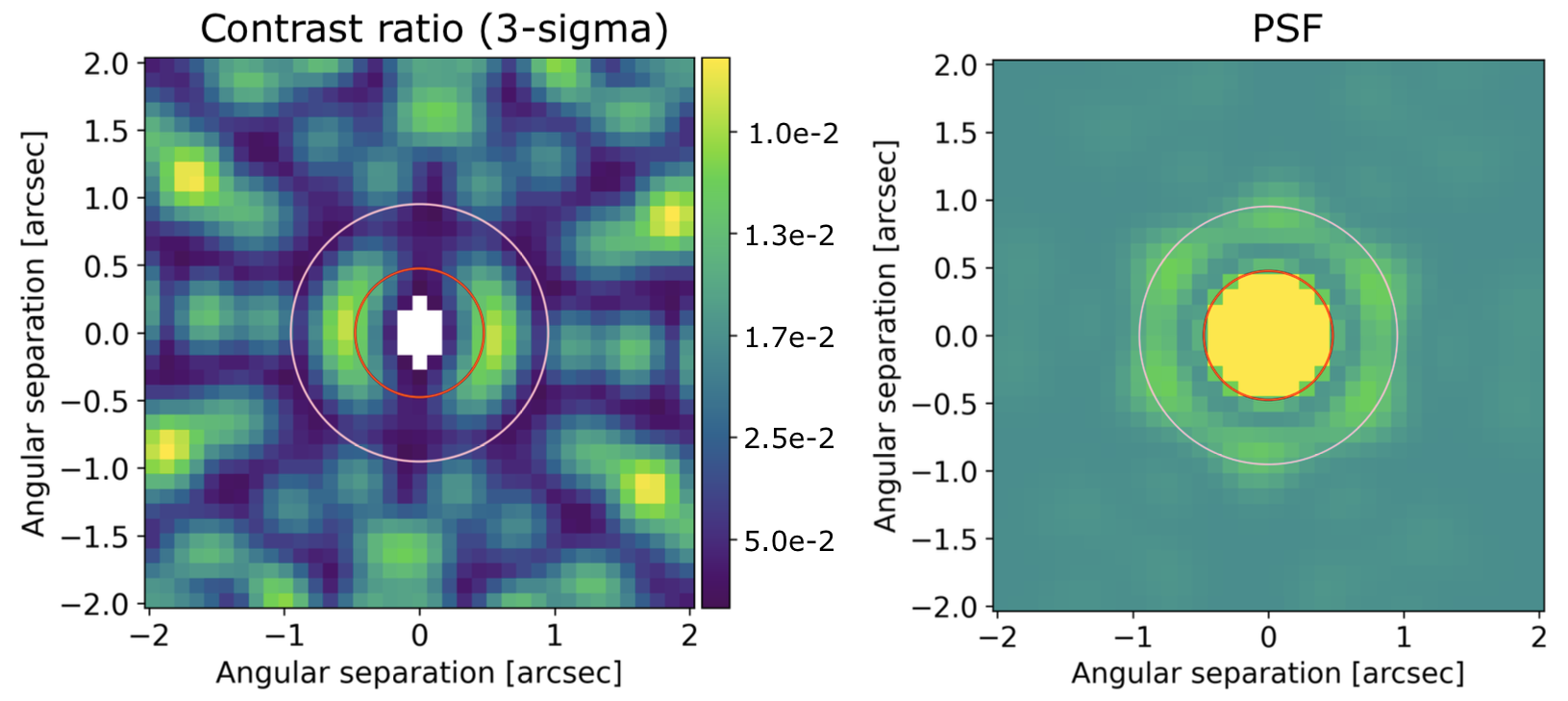}
    \caption{(left) Detection contrast map with KPI for WD2149+021 in the F1500W filter. Light regions correspond to high contrast sensitivity at 3-$\sigma$ confidence. The red and pink circles represent the 1\,$\lambda$/D and 2\,$\lambda$/D separations. (right) Representative point spread function in the same filter.  The best contrasts (in yellow in the left plot, contrast ratio of $\sim$0.01) are achieved at the separation corresponding to the first minimum of the Airy pattern, $\sim 1.22\, \lambda$/D. The other 3 WDs show similar maps, just scaled differently. The contrasts correspond to a 3$\sigma$ detection in the kernel phase space where binary model kernels are 3 times larger than the standard deviation of the measured kernels.}
    \label{fig:kpi}
\end{figure}

A pupil model from {\em jwst-kpi} \citep{kammerer.2023} built of 540 sub apertures distributed over 3 hexagonally-shaped rings for each primary mirror segment was adopted to sample interferometric baselines. The model optimized for MIRI is tilted by 4.84 degrees to account for the MIRI aperture position angle and only retains pupil points yielding a minimum of 50 redundant baselines. KPI analysis was performed on cropped images of 51$\times$51 pixels, or approximately 5\farcs6$\times$5\farcs6, keeping in mind that KPI's sensitivity advantage is limited to 0.5\,$\lambda$/D (0.25\,\arcsec) $\leq$ separation $\leq$ $\sim$ 2\,$\lambda$/D (1.0\,\arcsec). Figure~\ref{fig:kpi} presents the 3$\sigma$ contrast map produced in the KPI analysis of WD2149+021. For the 3 other targets, the map is very similar but scaled to different contrast ratios which correlate with the achieved point-source signal-to-noise.

\section{Resolved Planetary Companions}
\label{sec:resolved}

Previous work has already investigated possible resolved companions around three of the WDs considered here, with high probability of $p>98\%$ \citep{poulsen24, mullally24}. WD~2149+021 showed no high probability candidates, while WD~2105-820 and WD~1202-232 had high probability candidates. In this section we summarize the final contrast limits for each of our targets from a combination of direct imaging and KPI as well as report likely planet candidates within a projected separation of 100~au for each target.

\begin{figure*}
\centering
\includegraphics{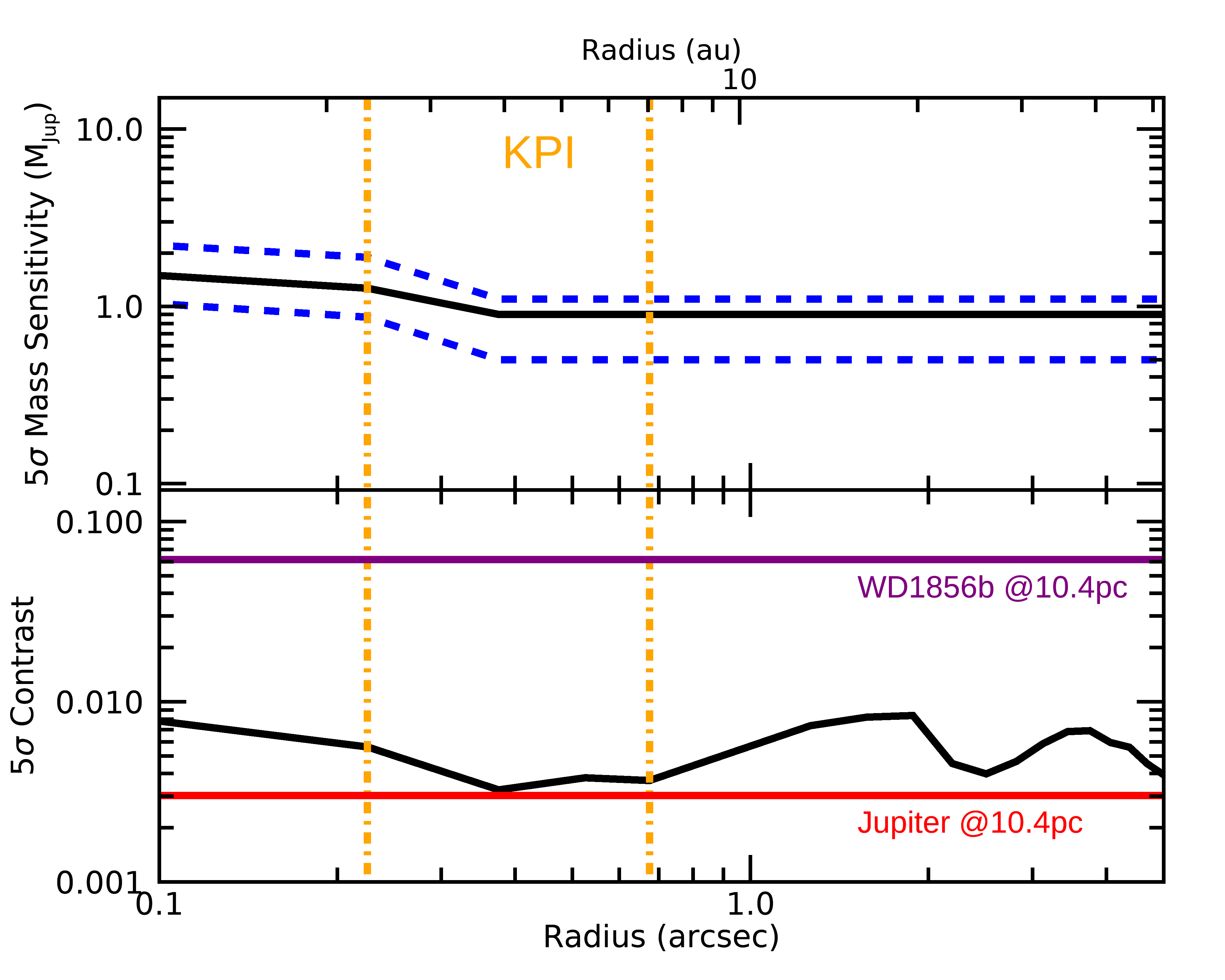}
\caption{\label{fig:contrast1202} (top) Planet sensitivity curve for WD~1202-232. The black curve represents the 5-$\sigma$ mass sensitivity limit in F1500W for a combination of KPI analysis ($<$0\farcs5) and with direct imaging with RDI. The flux ratio contrast is converted into mass sensitivity via the calculated isochrones as described in Section \ref{sec:resolved}. The blue dashed lines show the minimum and maximum masses assuming the age uncertainty of WD~1202-232. The lowest mass allowed by the models is 0.5~\Mjup, but the contrast sensitivity is a factor of 10 fainter than the predicted flux for 0.5~\Mjup.
(bottom) Contrast sensitivity of the observations represented by the black curve. Overplotted is the expected 15~\micron\ flux ratio of WD1856+584b and Jupiter at 10.4~pc.}

\end{figure*}

\begin{figure*}
\centering
\includegraphics{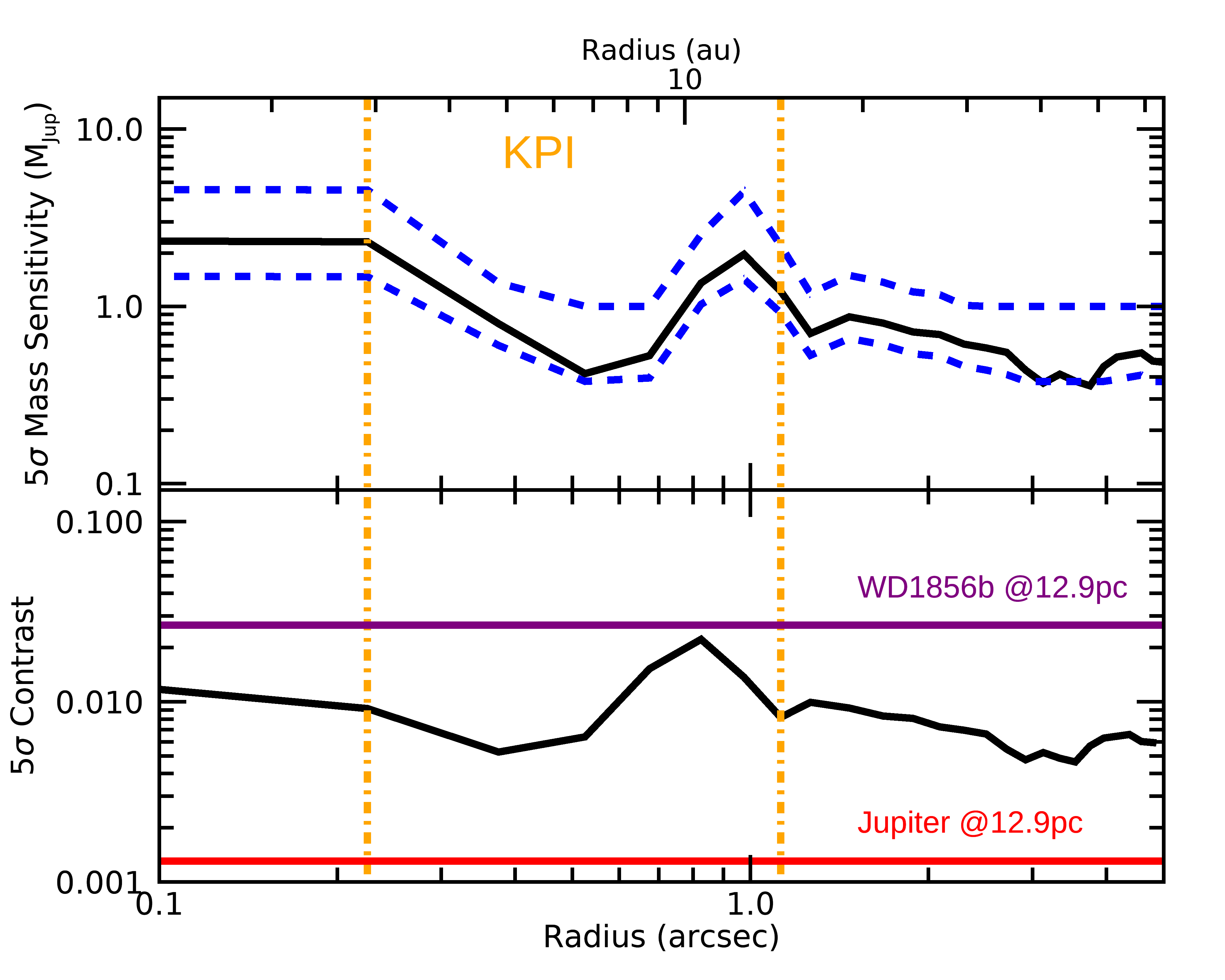}
\caption{\label{fig:contrast1620} (top) Planet sensitivity curve for WD~1620-391. The black curve represents the 5-$\sigma$ mass sensitivity limit in F1500W for a combination of KPI analysis ($<$0\farcs5) and with direct imaging with RDI. The flux ratio contrast is converted into mass sensitivity via the calculated isochrones as described in Section \ref{sec:resolved}. The blue dashed lines show the minimum and maximum masses assuming the age uncertainty of WD~1620-391. The lowest mass allowed by the models is 0.4~\Mjup and the contrast sensitivities are about at that level.
(bottom) Contrast sensitivity of the observations represented by the black curve. Overplotted is the expected 15~\micron\ flux ratio of WD1856+584b and Jupiter at 12.9~pc.}

\end{figure*}

\begin{figure*}
\centering
\includegraphics{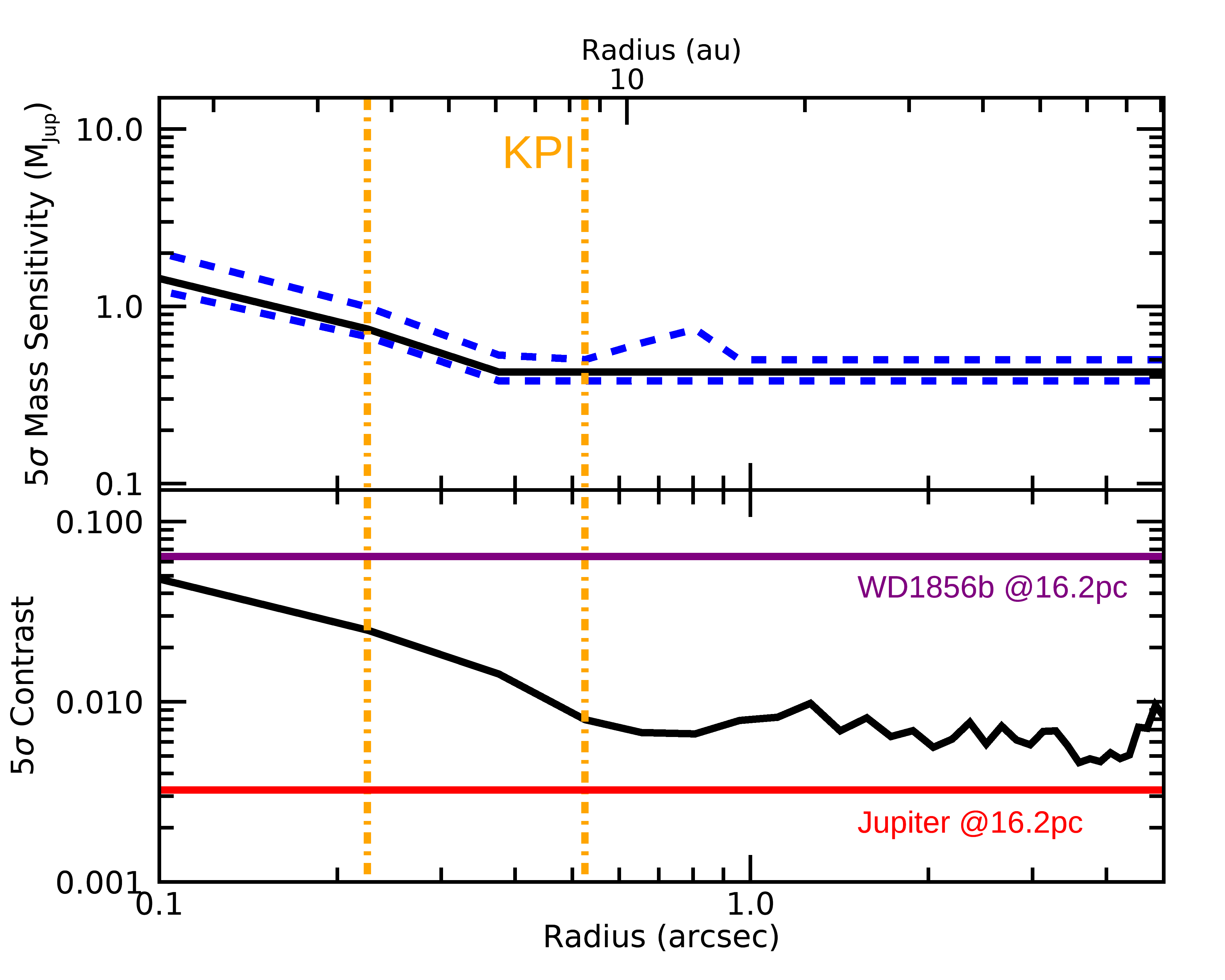}

\caption{\label{fig:contrast2105} (top) Planet sensitivity curve for WD~2105-820. The black curve represents the 5-$\sigma$ mass sensitivity limit in F1500W for a combination of KPI analysis ($<$0\farcs5) and with direct imaging with RDI. The flux ratio contrast is converted into mass sensitivity via the calculated isochrones as described in Section \ref{sec:resolved}. The blue dashed lines show the minimum and maximum masses assuming the age uncertainty of WD~2105-820. The lowest mass allowed by the models is 0.4~\Mjup, but the contrast sensitivity is a factor of 3 lower than the flux predicted for 0.4~\Mjup.
(bottom) Contrast sensitivity of the observations represented by the black curve. Overplotted is the expected 15~\micron\ flux ratio of WD1856+584b and Jupiter at 16.2~pc.}
\end{figure*}

\begin{figure*}
\centering
\includegraphics{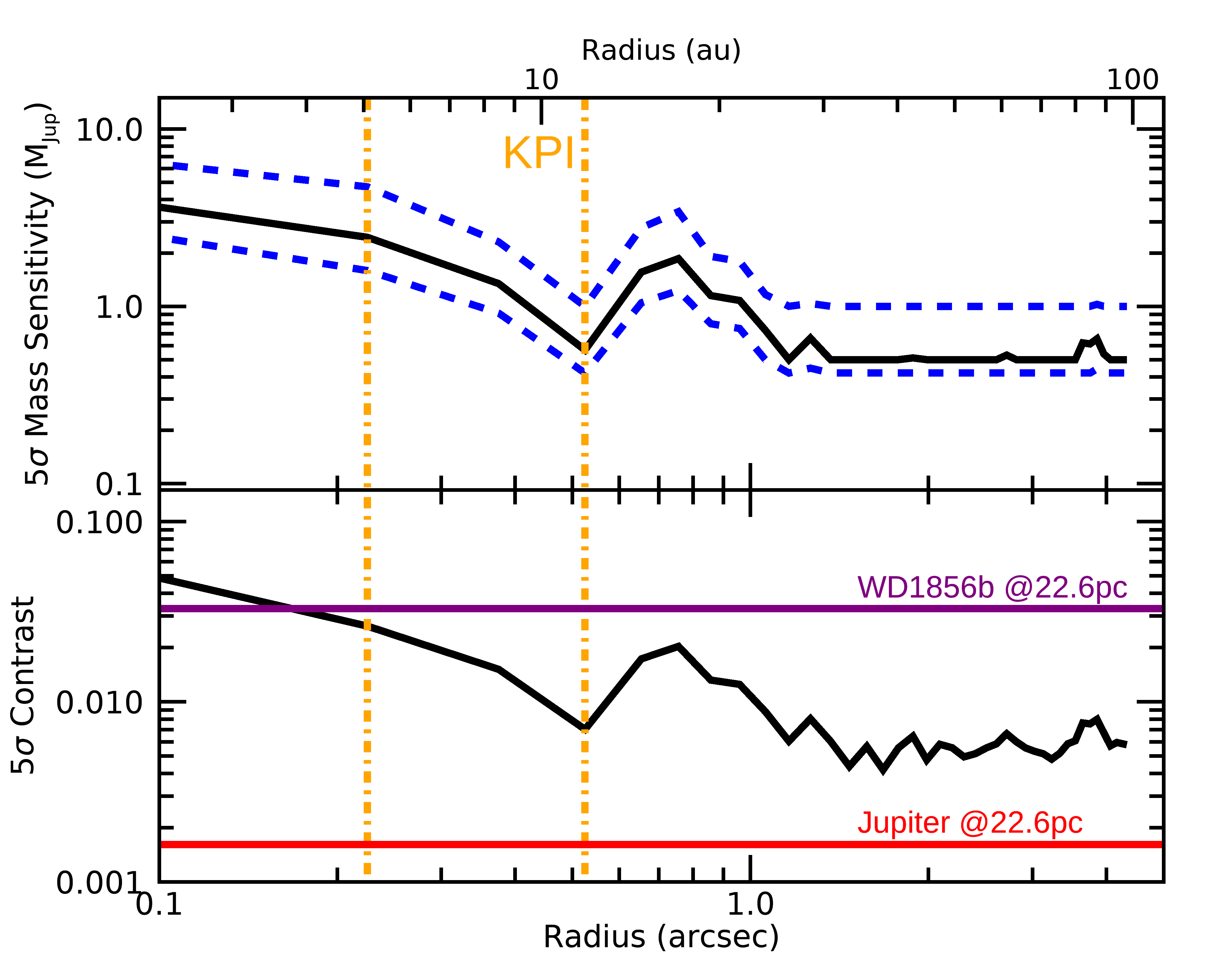}
\caption{\label{fig:contrast2149} (top) Planet sensitivity curve for WD~2149+021. The black curve represents the 5-$\sigma$ mass sensitivity limit in F1500W for a combination of KPI analysis ($<$0\farcs5) and with direct imaging with RDI. The flux ratio contrast is converted into mass sensitivity via the calculated isochrones as described in Section \ref{sec:resolved}. The blue dashed lines show the minimum and maximum masses assuming the age uncertainty of WD~2149+021. The lowest mass allowed by the models is 0.4~\Mjup, but the contrast sensitivity is 1.5 times fainter than the predicted flux for 0.4~\Mjup.
(bottom) Contrast sensitivity of the observations represented by the black curve. Overplotted is the expected 15~\micron\ flux ratio of WD1856+584b and Jupiter at 22.6~pc.}
\end{figure*}

Figures \ref{fig:contrast1202} to \ref{fig:contrast2149} show our 1-D contrast limits through a combination of our direct imaging and KPI techniques. The cross-over for both techniques is at approximately 0\farcs5, roughly at the spatial resolution limit for the F1500W filter. We convert the limiting flux ratios at F1500W, where we are most sensitive to resolved planetary companions, into masses based on the assumed mass, age, and distance to each target WD. We compare the contrast ratios to the recent detection of WD~1858+584b in F1500W (\Teff=184~K; M=5~\Mjup) \citep{limbach25} and Jupiter's expected 15~\micron\ flux (\Teff$\sim$150~K) scaled to the distance of each WD, which suggests our direct imaging with JWST/MIRI is sensitive to extremely cold giant planets even if our exact mass sensitivity is uncertain.

\begin{figure}
\plotone{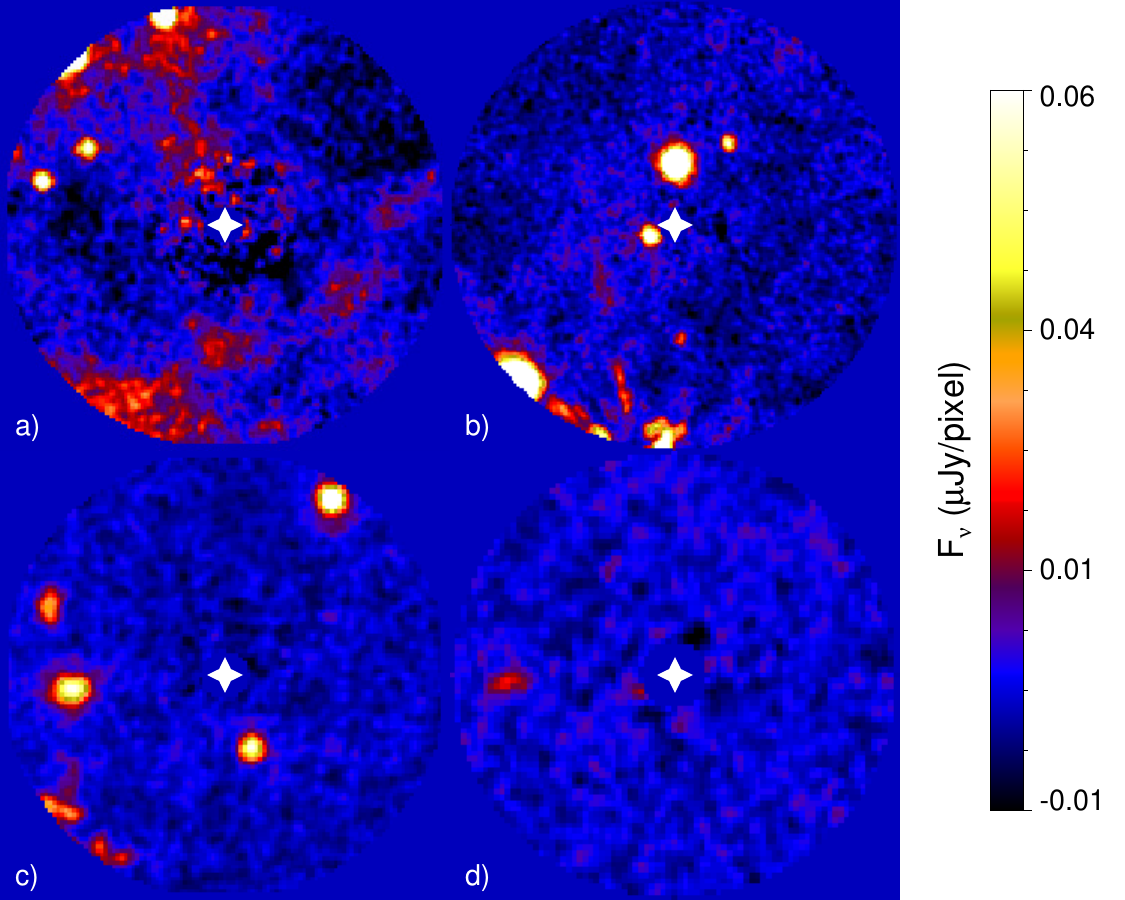}
\caption{\label{fig:candidates} PSF subtracted images of the four WD targets in F1500W. Each panel shows a circular search area of 100~au at the distance of the WD. The orientation of each panel is North up, East left. a) WD~1620-391: Galactic cirrus is present in the background of this image due to its low galactic latitude. Despite this, there are few mid-IR bright objects in the search area. The two point sources to the northeast are consistent with background stars. b) WD~1202-232: three primary sources besides the WD are present. The closest, a point source was reported in \citet{mullally24} as a planetary candidate. The others are consistent with background sources: one is spatially resolved and the other is blue across the MIRI filters. c) WD~2105-820: the only point source within a projected separation of 100~au is the planetary candidate to the southwest reported in \citet{mullally24}. The remaining sources are spatially resolved. d) WD~2149+021: The only significant source within a projected separation of 100~au is a spatially resolved faint galaxy in the background to the East.}
\end{figure}

We searched each of the four targets for resolved point sources within a projected separation of 100~au (See Figure \ref{fig:candidates}). We limit our search to this radius primarily because the probability of a coincident background point source with a red color (i.e. F1500W/F560W $>$ 1) reaches close to unity beyond this projected separation \citep{poulsen24,mullally24}.

Our selection criteria for candidates were well detected sources at F1500W that possessed FWHMs consistent with an unresolved point source and an F1500W/F560W ratio $>$1. With these criteria we recover the two planet candidates reported in \citet{mullally24}, but no other viable candidates.

\subsection{Limits to Resolved Planetary companions}
We calculated contrast curves in F1500W, the deepest images obtained for each WD, for the PSF subtracted images following the same procedures detailed in \citet{poulsen24}. For the given total age of each WD (t$_{cool}$+t$_{MS}$, we used cloudless Sonora-Bobcat grid models \citep{SonoraBob} above 200~K and cloudless Helios grid models \citep{linder19} below 2~\Mjup. We first calculated the proper \Teff, \logg\ combinations for a given age corresponding to various substellar companion masses from the Sonora-Bobcat grid,and then computed them for the youngest and oldest age as defined by the uncertainties in Table \ref{tab:wdparams}.

We then interpolated JWST fluxes from the grids of predicted fluxes for ages between 1.5-10 Gyr. For the Helios grid, fluxes for specific ages for 0.3, 0.5, 1, and 2 \Mjup\ were pre-calculated. We then converted the predicted fluxes or magnitudes to apparent fluxes accounting for the distance to each WD.

\section{Limits to Unresolved Planetary Companions}
\label{sec:unresolved}
For WD~1202-232 and WD~1620-391, the lack of any IR excess out to 21~\micron\ provides a strict limit to any unresolved planetary companions. For these targets we took the 5~$\sigma$ upper limit in flux at F2100W and compared these to the same planet isochrones as in Section \ref{sec:resolved} spanning the age uncertainties. For WD~2149+021 and WD~2105-820 we instead use the 3~$\sigma$ upper limit in flux at F1500W to estimate limits to unresolved planets under the assumption that the excess at F2100W is due instead to dust. For WD~2149+021 we find an upper mass limit of 4.2$^{+2.9}_{-1.3}$~\Mjup, and 1.8$^{+0.8}_{-0.3}$~\Mjup\ for WD~2105-820. For the remaining two WDs, we find an upper mass limit of 2.4$^{+1}_{-0.6}$~\Mjup\ for WD~1202-232 and 2.4$^{+2.2}_{-0.9}$~\Mjup\ for WD~1620-391. 


\section{Interpretation of Infrared Excesses Around WD~2149+021 and WD~2105-820} \label{sec:excessinterp}

The presence of $\sim$5-$\sigma$ excesses at F2100W for these two objects implies three likely possibilities: a) low luminosity dust, b) a cool companion radiating thermally, or c) an irradiated companion close to the WD. However, given the fact that we have a singular photometric excess point, it is difficult to find a unique model or answer. We thus work through the existing data to understand what is favored or disfavored for each scenario.

Prior to determining more complex situations, we instead calculate the available physical constraints on the excesses under the assumption of blackbody emission \citep{farihi14,poulsen24}. We calculated blackbody emission at a range of temperatures and total emitting areas and determined the hottest and coolest temperatures that succesfully matched the 21~\micron\ excess flux to within 1-$\sigma$ and did not exceed the 3-$\sigma$ upper limit excess at 15~\micron\ for WD~2149+021. For WD~2105-820 with a 2-$\sigma$ excess at 15~\micron, we calculated the maximum and minimum temperature/area combinations that matched to within 1-$\sigma$ of both the 15 and 21~\micron\ excesses for WD~2105-820. WD~2105-820's IR excess is consistent with a blackbody with T$_{\mathrm eff}$ between 170-280~K and with respective emitting areas between 7$\times$10$^{19}$~cm$^2$ and 1$\times$10$^{19}$~cm$^2$ (0.7 and 0.3~\Rjup). WD~2149+021's excesses are consistent with T$_{\mathrm eff}<330$ K and emitting area of $>$1.8$\times$10$^{19}$~cm$^2$ ($>$0.25~\Rjup). The implied radii of the excesses are tantalizingly similar to the range of radii seen in the Solar System giant planets and are at least an order of magnitude lower than that seen for typical warm WD debris disks.

\subsection{Cool or Low Luminosity Dust: Possible Disk Structures}
We first consider whether the IR excess could be caused by low luminosity versions of the dust disks seen around other WDs, something hitherto only proposed theoretically  before JWST \citep{bonsor17}. Typically WDs that show IR excesses between 5-10~\micron have high accretion rates--the typical accretion rates of our targets are about 3 orders of magnitude lower than that seen for a bright disk such as that seen around G29-38 \citep{xu14}. In terms of 21~\micron\ luminosity, G29-38 is $\approx$300-500 times brighter than WD~2149+021 and WD~2105-820. Thus, we explore if low luminosity and/or low mass versions of a G29-38 analog disk can match the data such that no excess is seen at wavelengths $<$10~\micron, but a significant excess is still present beyond 15~\micron.

To test the theory that the observed excess MIR emission is due to a dust disk rather than planet, we generate dust disk models and compare disk emission to the observed excesses. Since spatially resolving a white dwarf dust disk is not possible given the size of the disk, the exact structure of a white dwarf dust disk is uncertain, and must be inferred from disk thermal emission. Here, we assume three simple, distinct disk models, differentiated by the inner disk radius, and attempt to describe the observed excesses at WD 2105-820 and WD 2149+021 using each model. 

Disk thermal emission models are generated with radiative transfer code MCFOST \citep{mcfost2,mcfost1}. Due to a sparsity in data, we are limited to a "by-eye" modeling procedure, where the goodness of fit of a model is determined by if the model describes all photometric detections/upper limits. Significant degeneracy in dust disk parameter combinations producing the same flux at 21 $\mu$m leads to wide uncertainties on disk properties. If we however restrict ourselves to a fixed inner radius for each of the three models and vary only the mass of the dust in the disk, we are able to place estimates on how much dust would be required at different radial distances to produce the observed emission. Since we are primarily trying to determine if a dust disk is a plausible explanation for the excesses we observe, an extensive search of parameter space is outside the scope of this paper.

\begin{figure*} \label{fig:Debrisdisks}
\centering
\includegraphics[scale=0.8]{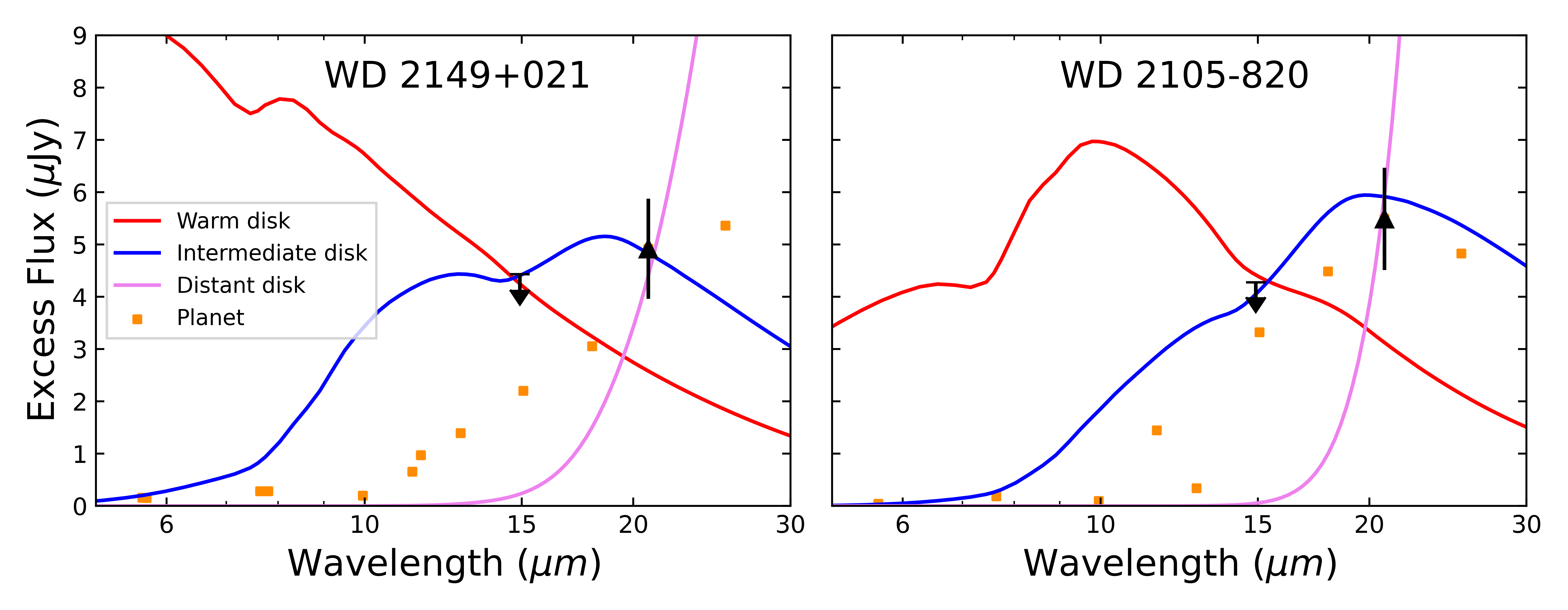}
\caption{Three different dust disk models fit to observed excess flux at WD 2149+021 and WD 2105-820 after a WD model is subtracted. Black points represent the observed photometric excesses at F2100W, and arrows represent upper detection limits for F1500W for both targets. The planet model fit to WD 2149+021 represents a 202K planet, and the model fit to WD 2105-820 represents a 190K planet.}

\label{fig:WD_SED}
\end{figure*}
\subsubsection{Warm dust disk near the Roche limit}

To develop a standard white dwarf dust disk model, we reproduce the approach of modeling the bright dust disk orbiting G29-38 presented in \citet{ballering22} using MCFOST. Once we generated a disk model that sufficiently fit observations of the G29-38 system, we altered the model to fit the much fainter disks around WD 2105-820 and WD 2149+021.

Disk composition is inferred to be similar to  the composition of material accreting onto the photosheres of polluted white dwarfs, which is primarily silicate dominated, bulk Earth material. \citet{reach09} finds strong evidence for amorphous carbon, amorphous/crystalline silicates, water ice, and metal sulfides in the material surrounding G29-38 using a detailed modeling approach. When modeling G29-38, \citet{ballering22} approximates dust composition using astronomical amorphous silicate optical constants \citep{draine84}, which reproduce the prominent silicate feature. Given that our goal is to determine if a disk is a plausible explanation for the observed excess rather than intricately model disk emission, we use the same constants as \citet{ballering22} for consistency between models.

We set the minimum and maximum particle sizes participating in emission at $<4$ $\mu m$ and $10^4$ $\mu m$, respectively. The distribution of particle sizes is described by a -3.5 power law \citep{kenyon17}, where smaller particles are more numerous, yet most mass resides in larger dust. Minimum grain size is assumed to be similar to that modeled around G29-38 in \citet{ballering22}, and maximum grain size is set to be similar to other dusty systems. To account for the mass of all particles in the disk, particularly those not participating in emission, we extrapolate the maximum particle size out to 300 km sized bodies for each disk model using the following: 

\[M_{disk} = M_{disk} \sqrt{\frac{a_{max, desired}}{a_{max, current}}}\]

White dwarf dust disk location is hypothesized to be set by the sublimation radius and Roche limit of the white dwarf. However, \citet{ballering22} finds a more distant disk, with the inner radius existing just exterior to the theoretical Roche limit of the white dwarf \citep{jura03}. The likely explanation for the location of the inner radius is that the disk's vertical scale height is larger than the white dwarf's radius, so the bulk of disk emission is from the hot, inner edge of the disk and does not necessarily require significant material closer to the white dwarf. Since the white dwarfs in our sample have masses and radii similar to G29-28, and thus similar Roche radii, we choose inner and outer disk locations close to each white dwarf's Roche radius as was modeled by \citet{ballering22}. A white dwarf dust disk outer radius of $100-200 R_{WD}$ has been observed/modeled in \citet{steele21} and directly observed with Doppler tomography in \citet{manser16}, validating our choice of disk location.

Dust mass is determined after considering the constraints on other parameters, and we find a narrow range of masses accurately models the data. We fix the gas to dust ratio at 0.1, given that \citet{ballering22} did not model gas around G29-38, and we found that the gas to dust ratio did not significantly affect the disk spectrum. We also found no constraint on disk inclination, as our disk models are primarily optically thin and therefore emission does not depend on disk orientation.  

\subsubsection{Intermediate disk}

For each target, we additionally produce an intermediate disk model, with an inner radius located around 10 times further from the star than the warm disk model described above. This model was included because a similar dust model can be fit to excesses observed at both 10 and 15 $\mu m$ in Poulsen et al. 2025 (in prep). For this disk, we assume 10\% disk width and a particle size distribution similar to the \citet{ballering22} model. However, there is no known scenario that would produce dust in a white dwarf system interior to one AU yet exterior to the white dwarf's Roche limit. Via \citet{jura08}, material interior to 1 AU with radii smaller than 1-10 km will be removed from the system during stellar evolution, so micron-sized dust in a disk at this location would need to have formed after stellar evolution. 

\subsubsection{Distant, massive disk}

We also explore the possibility that dust is located in a cold, distant disk between 3 - 3.3 AU, similar to our solar system's asteroid belt. \citet{jura08} describes a white dwarf pollution scenario in which the material that forms a dust disk and pollutes the white dwarf photosphere originate in an asteroid belt residual from the progenitor star, altered by the host star's red giant branch evolution. For $1-2 M_{\odot}$ progenitor stars such as those in our sample, \citet{jura08} predicts asteroids larger than $1-10$ km will remain in the disk, while smaller particles will be removed. Such disks would require on the order of $10^{25}$ g of material, slightly more massive than the solar system disk with a mass of $1.8 \times 10^{24}$ g. 

However, no such disks have been observed around white dwarfs, but this may also be due to a lack of sensitivity to such disks prior to JWST. The only such effort was to detect a solar-system-like asteroid belt around white dwarf G29-38 with a known debris disk \citep{farihi14}, which resulted in a non-detection and upper mass limits of $10^{23}$ g interior to $11$ AU. Therefore, we leave this distant disk as a hypothetical, similar to our Solar System's asteroid belt. 

\subsection{Comparison of Low Luminosity Disks with 21~\micron excesses}

\begin{table*}[]
\centering
\begin{tabular}{l|lrr}
Disk model               & Disk location (AU) & \multicolumn{1}{l}{$M_{disk}$ (g) } & Particle sizes ($\mu m$)             \\ \hline \hline
\multicolumn{4}{c}{\textbf{WD 2105-820}}                                                                                      \\ \hline
Warm disk at Roche limit & 0.0055 - 0.006     & $1.5 \times 10^{17}$                   & $1.4 - 10^{4}$ \\
Intermediate disk        & 0.05-0.055         & $3.4 \times 10^{18}$                      & $1 - 10^{4}$    \\
Distant disk             & 3 - 3.3            & $2.5 \times 10^{23}$                       & $1 - 10^{4}$     \\ \hline
\multicolumn{4}{c}{\textbf{WD 2149+021}}                                                                \\ \hline
Warm disk at Roche limit & 0.0055 - 0.006     & $1.5 \times 10^{17}$                   & $1.4 - 10^{4}$ \\
Intermediate disk        & 0.05-0.055         & $2.3 \times 10^{18}$                      & $1 - 10^{4}$    \\
Distant disk             & 3 - 3.3            & $4 \times 10^{21}$                       & $1 - 10^{4}$     
\end{tabular}
\caption{Parameters of three different disk models fit to excesses around both white dwarfs in this study. }
\end{table*}

\subsubsection{WD 2149+021}

WD 2149+021's excess is reasonably described by the distant disk model, and the warm dust disk model is less than $3 \sigma$ ($2.76 \sigma$) from the excess at $21 \mu m$. An intermediate disk model describes the excess while remaining under the 15 $\mu m$ detection limit. The distant disk model requires on the order of $10^{21}$ g of material, which is three orders of magnitude less massive than our solar system's asteroid belt \citep{binzel00}. A 203 K planet model accurately describes both the excess and falls below the $15 \mu m$ upper limit. Therefore, no disk model can be entirely ruled out.

\subsubsection{WD 2105-820}
The most plausible warm disk model lays at just over $3 \sigma$ from the $21 \mu$m excess, making it a slightly less likely explanation for the excess. The intermediate disk can be made to describe both constraints. The distant disk model requires on the order of $10^{23}$ g of mass. However, this represents the mass of particles between 1 - $10^{4}$ $\mu m$, and a distant disk would be expected to contain much larger bodies, which would significantly increase the mass of the disk.
N-body simulations of planetary systems perturbing asteroid belts often require more mass than that of the Solar System to explain the high accretion rates seen around other WDs \citep{debes12}. A planet model of 190 K can be fit to the 21 $\mu m$ excesses within $1 \sigma$.  

\subsubsection{Which Disks are Most Plausible?}
 
 A plausible dust disk with the same structure and similar composition to brighter dusty white dwarf disks, such as that observed around G29-39 \citep{ballering22}, can somewhat describe excess MIR emission around both WD 2105-820 and WD 2149+021 if we infer less disk mass. Similarly, a model where a larger mass of dust resides around 3 AU can also be made to fit the excesses, with dust masses between 1 - $10^{4}$ $\mu m$ grain sizes similar to the mass of the Solar System's asteroid belt. Additionally, the mass of the Eps Eri Belt \citep{krishnan} is hypothesized at around $10^{24}$ g, so we may not rule out the possibility that a further out dust disk causing the excess without more data. While all three kinds of dust disks considered here are somewhat consistent with the data, the intermediate and cold dust disks which best fit the data are not well physically motivated since the RGB and AGB phases of stellar evolution should directly remove most minor bodies from orbital separations $<$3~au. We thus argue that planetary mass companions are more likely to explain the 21~\micron\ excesses we detect.
 
 Given that reasonable dust disk models can be derived from a faint MIR excess, we demonstrate that JWST has the capabilities to detect and characterize a wider swath of dust disks around white dwarfs than was historically possible with WISE and Spitzer. From disk SED modeling, we predict around $>$6\% disk emission at 10 microns in the scenario of warm dust located close to the star, so a spectrum spanning 5-13 $\mu$m taken with the MIRI Low-Resolution Spectrograph (MIRI-LRS) would provide more definitive evidence for the presence of warm dust disks around our targets.

\subsection{Thermal Emission from Planets}

Similar to dust disks, the presence of an excess at one wavelength can be compared to cool exoplanet atmosphere models to provide a constraint on the possible mass of a planetary companion. This and constraints on how close the planet is to the WD can be used to test models for planetary evolution during post-main sequence evolution. Currently, it is plausible that most Jovian planets survive beyond a given radius \citep{mustill12}, with a smaller fraction of planets that might be found interior, and virtually no planets interior to this survival radius. Given the small number of targets, it is striking that two WDs show some evidence of a planet being present, with a possibility that WD 2105-820 has both an interior planet and one at 34~au \citep{mullally24}.

Based on the distance and 21~\micron\ flux, we compare to the same grid of models used to determine our sensitivity to resolved planetary companions. The excess flux for WD~2149+021, assuming a total system age of 3~Gyr, is consistent with a $3^{+3}_{-2}$~\Mjup\ companion and a \Teff=203$\pm$16~K. While we quote uncertainties driven by the system age uncertainties, we note that our inferences are heavily dependent on the accuracy of our models, which are largely untested at these temperatures. WD~2105-820's excess is consistent with a 1.4$^{+0.4}_{-0.2}$~\Mjup\ companion with \Teff=171$\pm$10~K. If confirmed, these would be the some of the coolest planets yet detected and similar to the recently announced cool planet candidate in orbit around WD~0310-668 (\Teff=248~K) \citep{limbach24} and the transiting planet orbiting WD 1856+534b (\Teff=186~K)\citep{limbach25}.

A combination of our PSF subtraction and KPI limits for point sources constrains the location of the excesses to be within 4 and 8~au for WD~2105-820 and WD~2149+021 respectively. If we include similar limits from GAIA astrometric noise based on the fact that both stars have renormalized unit weight errors (RUWE) of less than 1.4, the planets would be constrained to reside interior to 2~au \citep[see][]{limbach24}.


\subsection{Irradiated Planets}

As explored in \citet{poulsen24} and references therein, close-in exoplanets will also show an IR excess and radiate hotter than would be expected for a given system age. Given the inferred range of temperatures due to these excesses we can also infer the range of orbital radii such planets would reside by estimating an equilibrium temperature based on our blackbody estimates (T$_{\mathrm eq}$) \citep{poulsen24}:

\begin{equation}
    T_{\mathrm {eq}}= T_{\mathrm{eff}}\left(1-\alpha\right)^{0.25}\sqrt{\frac{R_{\mathrm {WD}}}{2~a}}
\end{equation}

and solving for semi-major axis assuming a geometric albedo $\alpha$=0.3 and R$_{\mathrm WD}$=0.011~\Rsun\ for both WDs. Under these assumptions, our observations are consistent with irradiated planets residing at between 0.02-0.15~au for WD~2149+021 and 0.01-0.1~au for WD~2105-820, corresponding to orbital periods of 1.3-40~d and 0.5-15~d for WD~2149+021 and WD~2105-820,  respectively.

\section{Discussion} \label{sec:discussion}
Our study of four white dwarf stars has found intriguing evidence of planetary systems that may be driving the host star's pollution, thus elucidating the type of planetary architecture that most commonly drives the tidal disruption of rocky bodies in orbit around post-main sequence systems. Out of a sample of four polluted white dwarfs, a shocking three targets show evidence of either planets, or dust, or both. We summarize all candidate planets for our systems in Table \ref{tab:candidates}, with new mass estimates for the resolved planet candidates based on a best fit to the 21~\micron\ fluxes and accounting for age uncertainties. We choose the 21~\micron\ flux as the basis for our mass estimate to be consistent with the unresolved excesses. If we include the recent results from the MEOW survey \citep{limbach24}, a third metal-polluted white dwarf also shows an infrared excess similar to those seen with our targets, located $<$2~au from its host. Other planetary mass objects with similar low temperatures to our resolved candidates and unresolved excesses are available for comparison with fluxes observed with JWST: WD 1865+534b \citep{limbach25}, a transiting planet at 25~pc with JWST MIRI detections, Eps Indi Bb at 3.6~pc \citep{matthews24}, and WD 0806-661b at 19.23~pc \citep{voyer25}.

\begin{deluxetable}{lcccc}
\label{tab:candidates}
\caption{Summary of candidate planet properties} 
\tablehead{
\colhead{WD} & \colhead{Unresolved M} & \colhead{Unresolved $a_{\mathrm proj}$} & \colhead{Resolved M} & \colhead{Resolved $a_{\mathrm proj}$} \\
\colhead{} & \colhead{(\Mjup)} & \colhead{(au)} & \colhead{(\Mjup)} & \colhead{(au)}
}
\startdata
WD~1202-232 & ... & ... & 1 & 11 \\
WD~2105-820 & 1.4$^{+0.4}_{-0.2}$ & $<$4 & 1.6$^{+0.4}_{-0.2}$ & 34 \\
WD~2149+021 & 3$^{+3}_{-2}$ & $<$8 & ... & ... \\
\enddata
\end{deluxetable}

It is interesting to note that one WD in our sample, WD~2105-820, posseses a directly imaged candidate planetary mass companion at $\sim$34~au and the presence of a significant IR excess consistent with a planetary mass companion at closer separation $<$4~au. Intriguingly, the 21~\micron\ flux of both the IR excess and the companion are equivalent within the uncertainties \citep{mullally24}. If both are confirmed to be planets, then this would represent the first multi-planet WD system to be discovered. The inferred masses for both objects then are close to Jupiter's and the difference in separations $\sim$30~au is significantly larger than that seen in our own Solar System. This could hint at the possibility of dynamical interactions that sent one of the giant planets into an eccentric orbit with a large semi-major axis \citep[e.g.][]{debes02}. If that were true, we would predict that both planets would have eccentric orbits. Gaia DR4 and future JWST MIRI imaging could provide constraints on the orbital parameters of the inner and outer planet respectively.

While it is premature to draw strong conclusions from this study due to small-number statistics, it is helpful to think about possible future trends our work implies, once the IR-excesses we detected are characterized and when the resolved candidates are confirmed or refuted as planetary companions. We thus consider the case where a) all candidates are confirmed as planetary companions, b) only IR-excesses are confirmed as planetary companions, or c) no planetary companions are confirmed and the IR-excesses are due to dust close to the WD.

In the first case, we would be faced with a situation where 75\% of our sample stars had at least 1 giant planet with a mass of $\sim$1-5\Mjup, at orbital separations that range from as close as a few tenths of an au to 34~au. Despite the small number uncertainties, it would still imply, based on binomial probability statistics for 4 detections in 4 trials, an occurrence rate of giant planets of $>$55\% at 95\% confidence. In this scenario giant planets are overrepresented compared to the occurrence rate of giant planets in main sequence star RV surveys \citep{fernandes19} and at a possible range of orbital radii that might directly challenge our notions of planetary survival, particularly if the IR-excess candidates reside at very close orbital separations. This outcome strongly favors the validity of the giant planet driven models of WD accretion \citep[e.g.]{debes02} and suggest that pollution is not largely driven by widely separated binaries \citep{Veras2018} or other scenarios that do not require a giant planet \citep[e.g.][]{Zuckermon1998, OConnor2022}. 

The second case, where only the infrared excess planets are confirmed, is not that different in result from the first case, with the exception that WD pollution may be driven by a population of giant planets fairly close in primordial orbital separation to Jupiter. With upper limits to orbital separation at 4 and 8~au for WD~2105-820 and WD~2149+021 respectively, the planets may be very close to the minimum survivable distance from the host star when the star was a red giant or asymptotic giant. \citet{mustill12}, using simulations, predicted the final orbital radii of planets that just escaped engulfment, for both circular and elliptical orbits as a function of initial stellar mass. Assuming the best estimates of initial mass for WD~2105-820 and WD~2149+021 (2.5 and 1.5\Msun, respectively), the outermost surviving planets should reside beyond $\sim$6 and 3.5~au under the assumption that the initial planet orbits were circular. This puts the WD~2105-820 excess at odds with theoretical predictions which could imply that the planet had a non-circular orbit or that tidal evolution of exoplanets is stronger than predicted.

In the third case where none of the planets are confirmed, if our sample is representative, it would suggest that a large fraction of polluted white dwarfs have cool IR excesses, even at accretion rates that are orders of magnitude smaller than those seen for existing warm WD disks detected at shorter wavelengths. This scenario would then favor a situation where giant planets larger than 1\Mjup\ are not the dominant perturber of minor bodies. It is possible that undetectable lower mass giant planets cause this pollution. Depending on mass, this scenario would most likely require a significant number of lower mass planets to drive the metal pollution \citep{OConnor2022}. 

The results of our exploratory program demonstrate the revolutionary sensitivity that JWST/MIRI has for detecting planetary systems around nearby WDs and represents an opportunity for the future exploration of dead planetary systems, especially for polluted WDs. A search of a larger number of polluted WDs within 25~pc could yield a trove of interesting mid-IR detections and directly imaged planets. One problem is that our knowledge of metal line WDs is incomplete--out of a total of 58 stars within 25~pc and with \Teff$>$9000~K, only 5, or 10\%, are known to contain metals, lower than that seen in other samples of warm WDs. This is partially due to new WDs being discovered by {\em Gaia}, and a lack of high spectral resolution spectra in either the visible or the UV. Ideally, a volume limited search for metals could be conducted to help JWST focus in on the highest value targets. 

Moving forward, it is clear that the faintest IR-excesses will be hard to distinguish between plausible debris disks and planetary companions even with JWST. They are sufficiently faint that it is nearly impossible to get a direct spectrum beyond 15~\micron\, and thus will require sensitive searches for silicate emission lines at 10~\micron\ with MIRI/LRS or additional photometry at 18 and 25~\micron\ in the F1800W and F2500W filters with the MIRI imager \citep[e.g.][]{limbach24}. Given the lower sensitivity of the F2500W filter, where most cool planet SEDs are peaking and dust SEDs from warmer disks are expected to be getting fainter some of these excesses will have to wait for future observatories sensitive to $\sim\mu$Jy sources beyond 25~\micron, such as the proposed PRIMA mission.

\begin{acknowledgments}
This work is based on observations made with the NASA/ESA/CSA James Webb Space Telescope which is operated by the Association of Universities for Research in Astronomy, Inc., under NASA contract NAS 5-03127 for JWST. These observations are associated with program GO1911. Some of the data presented in this paper were obtained from the Mikulski Archive for Space Telescopes (MAST) at the Space Telescope Science Institute. The specific JWST/MIRI observations analyzed can be accessed via \dataset[https://doi.org/10.17909/k3zk-rb06]{https://doi.org/10.17909/k3zk-rb06}. STScI is operated by the Association of Universities for Research in Astronomy, Inc., under NASA contract NAS5–26555. Support to MAST for these data is provided by the NASA Office of Space Science via grant NAG5–7584 and by other grants and contracts. M.K. acknowledges support by the NSF under grant AST-2205736 and NASA under grant No. 80NSSC22K0479. L.A. acknowledges the support from the Canadian Space Agency through grant 22JWGO1-02.
This publication makes use of data products from the Wide-
field Infrared Survey Explorer, which is a joint project of the
University of California, Los Angeles, and the Jet Propulsion
Laboratory/California Institute of Technology, and NEO-
WISE, which is a project of the Jet Propulsion Laboratory/
California Institute of Technology. WISE and NEOWISE are
funded by the National Aeronautics and Space Administration.
\end{acknowledgments}

\bibliography{miriwd}{}
\bibliographystyle{aasjournal}


\end{document}